 \documentclass[prc,amsmath,amsfonts,showpacs,eqsecnum,nofootinbib]{revtex4-2}
\usepackage{graphicx}
\usepackage{amsmath}   
\usepackage{pstcol}    
\usepackage{bm}        

\begin{document}


\title{Topics in Kadyshevky Field Theory 
\protect\footnote{For the original version: see Report THEF-NIJM 09.07,\\
http://nn-online.org/eprints(2009).}                                 
}

\author{Th.A. Rijken}
\author{J.W. Wagenaar}
\affiliation{Institute for Mathematics, Astrophysics and Particle Physics, \\
 University of Nijmegen, Nijmegen, the Netherlands}

 \date{version of: \today}

\begin{abstract}
In these notes we deal with several field theoretical topics  in the 
framework of the Quantum Field Theory as developed by Kadyshevsky. 
The main motivation for studying the Kadyshevsky formalism is that  
in the Kadyshevsky-graphs, in contrast with the Feynman graphs, 
the particles remain on-mass-shell. This facilitates the use of phenomenological 
form factors, {\it e.g.} Gaussian ones. \\
In the first part we first construct the second-quantized quasi-particle formalism, 
which is employed to develop the functional integral formalism. In the latter
we develop the path-integral, the Schwinger-Symanzik equations, generalized
Wightman functions, and the Kadyshevky reduction formulas.
In the second part we cover the topics: 
Relation between the Feynman and Kadyshevsky perturbation theory,    
and the Gross-Jackiw method in the Kadyshevsky formalism. 
The latter method is applied to the Kadyshevsky formalism for interaction Lagrangians 
	with derivatives, in particularly for 
pion-nucleon interactions: (i) pseudo-vector $NN\pi$-, (ii) vector $NN\rho$-   
and (iii) gauge-invariant $\Delta_{33}N\pi$-coupling.
\end{abstract}
\pacs{13.75.Cs, 12.39.Pn, 21.30.+y}

\maketitle
 
\section{Introduction}                                          
\label{sec:1}
In these notes several field theoretical topics are treated in the 
framework of the Quantum Field Theory as developed by Kadyshevsky. 
Apart from the topics based on the functional integral formalism,
these topics have also been discussed in \cite{Wag09a,Wag09b,Wag09c}, 
in particularly the covariant formulation of (absolute) baryon-antibaryon 
pair-suppression on the level of baryons and mesons.
\\
The Kadyshevky-rules for covariant perturbation theory are
derived from the same standard formula for the S-matrix, which is used to derive
the Feynman rules. This demonstrates immediately the equivalence of the Kadyshevsky-
and Feynman-formalism.\\
For our main motivation for studying the Kadyshevsky formalism, we note that  
in the Kadyshevsky-graphs, in contrast with the Feynman graphs, 
the particles remain on-mass-shell.\\
Phenomenologically this can be exploited e.g. to introduce phenomenological
vertex form factors which suppress the transitions between the
positive and the negative energy solutions in a covariant way.
These kind of form factors
are easily handled in the Kadyshevsky-formalism, and can be shown
to be effective. This is impossible in the usual treatment
using Feynman graphs. Therefore, pair-suppression can be introduced
phenomenologically and covariantly, and is accessible for an analysis 
using a fit to the meson-nucleon data.
Also, Gaussian form factors, describing {\it e.g.} the internal quark structure of the nucleons,
can be used which are a function of the montum transfer Mandelstam variables. \\
An advantage of the Feynman formalism w.r.t. the Kadyshevsky formalism, likewise to the 
"old perturbation" formalism, is to have less number of diagrams. 
This is in particular important when
large order diagrams have to be calculated, as {\it e.g.} in higher-order corrections and 
multi-particle production. Nowadays this disadvantage is more and more unimportant 
since the availability of powerful algorithmen for automatic calculations.

The contents of this paper is arranged into seven sections and two appendices.
In section~\ref{app:Aa} the Kadyshevky formalism is reviewed and the relation and 
differences with the Feynman formalism are indicated.  
In section~\ref{app:B} we introduce the second quantization formalism for quasi-particles
in momentum space. This is e.g. a necessary step for the development of both the
path-integral formalism and the Kadyshevsky LSZ-type \cite{LSZ54} of reduction formulas.
In section~\ref{app:C} a functional integral formalism for the Kadyshevsky theory is
given, which leads for example to a path-integral formulation, 
Schwinger-Symanzik equations, the Kadyshevsky Wightman functions related to a 
generating functional. Also, we derive Kadyshevsky reduction formulas.\\
In section~\ref{app:T} the Kadyshevsky formalism for interactions with derivatives
is treated. Here we introduce the covariant T$^*$- and R$^*$-product in order to 
arrive at a {\it covariant} and {\it frame independent} S-matrix, both in the Feynman and
Kadyshevsky formalism. In section~\ref{app:EX} we apply the theory for several
relevant interaction Lagrangians.
In section~\ref{sec:Z} the paper is closed with Discussion and Conclusions.\\                      
In Appendix~\ref{app:A} the Kadyshevsky rules for graphs in momtum-space are listed.
In Appendix~\ref{app:L} Kadyshevsky $\widetilde{T}$-products and Wick's Theorem are given.
\onecolumngrid

\section{Kadyshevsky-basics in Configuration- and Momentum-Space}
\label{app:Aa}
In this section the Kadyshevsky formalism is briefly introduced by using the 
S-matrix formula in quantum-field-theory as a strarting point, and going from there to
the rules for the Kadyshevsky-diagrams \cite{Kad64,Kad67,Kad68,Kad70}.           
We follow the set up of the appendices B in \cite{BD65} where the
rules for the Feynman-graphs are given. The differences will then 
come to the surface in a most transparent manner.
Starting from the expression of the S-operator, one has \cite{Schweber,IZ80}
\begin{eqnarray}
 S &=& 1 + \sum_{n=1}^\infty \left(\frac{i}{\hbar}\right)^n
 \int_{-\infty}^{+\infty} \ldots \int_{-\infty}^{+\infty} 
\theta(x_{n}^0-x_{n-1}^0) \theta(x_{n-1}^0-x_{n-2}^0)
 \ldots \theta(x_{2}^0-x_{1}^0) 
\cdot \nonumber \\ && \hspace{2cm} \times 
{\cal L}_I(x_n)\ {\cal L}_{I}(x_{n-1}) \ldots {\cal L}_I(x_1) 
\cdot d^4x_n \ldots d^4x_1 \nonumber \\ 
 &\equiv& 1 + \sum_{n=1}^{\infty} S_n\ ,
\label{app:A.1}\end{eqnarray}
we follow \cite{Kad64} and introduce the time-like vector  
$n^\mu$ with $n^2=n_0^2-{\bf n}^2=1, n_0 > 0$. Then (\ref{app:A.1}) 
can be brought into a completely 4-dimensional covariant form, although 
 frame-dependent, by the replacement
\begin{equation}
 \theta(x^0) \rightarrow \theta( x\cdot n)\ \ ,\ \ 
 n\cdot x = n_0 x^0 - {\bf n}\cdot{\bf x}\ .
\label{app:A.2}\end{equation}
This gives ($ \hbar =1$)
\begin{eqnarray}
 S_n &=& i^n
 \int_{-\infty}^{+\infty} \ldots \int_{-\infty}^{+\infty} 
 \theta[n\cdot(x_n-x_{n-1})] \theta[n\cdot(x_{n-1}-x_{n-2})]
 \ldots \theta[n\cdot(x_{2}-x_{1})] 
 \cdot\nonumber\\ && \hspace{2cm} \times
 {\cal L}_I(x_n)\ {\cal L}_{I}(x_{n-1}) \ldots {\cal L}_I(x_1) 
\cdot d^4x_n \ldots d^4x_1\ .            
\label{app:A.3}\end{eqnarray}
The equivalence of $S_n$ in equations (\ref{app:A.1}) and (\ref{app:A.3}) can be
seen as follows. Assuming that the S-matrix defined in (\ref{app:A.1}) is 
Lorentz-invariant, and realizing that (\ref{app:A.1}) and (\ref{app:A.3}) are 
identical in the frame where $n^\mu = (1,{\bf 0})$, it follows that they are
equivalent in all frames because the expression in (\ref{app:A.3}) is manifest
Lorentz-invariant. Also, it follows that the S-matrix defined in (\ref{app:A.3})
is independent of the four-vector $n^\mu$. A more explicit elaboration on this
issue and others is given in appendix~\ref{app:L}.\\

\noindent From the expression (\ref{app:A.3}) one can work out the rules for
the Kadyshevsky graphs in a way which parallels the derivation of the 
Feynman rules. The differences come from the treatment of the 
$\theta$-functions. In the case of the Feynman graphs one includes the 
$\theta$-functions into the propagators by applying the Wick-expansion to
the $T$-products of the field operators. In the case of the Kadyshevsky
graphs one employs a four-dimensional form of the $\theta$-functions, 
 exploiting (\ref{app:A.2}),
\begin{equation}
 \theta(n\cdot x) = -\frac{1}{2\pi i}\int_{-\infty}^{+\infty} d\kappa
 \frac{\exp\left[-i\kappa(n\cdot x)\right]}{\kappa+i\epsilon}\ ,
\label{app:A.4}\end{equation}
and one applies the Wick-expansion to the ordinary products of the
field operators. Then, the propagators are given by
\begin{eqnarray}
	\langle 0|\phi(x)\phi(y)|0\rangle &=& \int
	\Delta^{(+)}(x-y; \mu^2) = \int
 \frac{d^4q}{(2\pi)^3}\theta(q_0)\delta(q^2-\mu^2)\ e^{-iq\cdot(x-y)}\nonumber\\
 \langle 0|A_\mu(x)A_\nu(y)|0\rangle &=& D^{(+)}_{\mu\nu}(x-y) = -g_{\mu\nu}
 \int\frac{d^4q}{(2\pi)^3}\theta(q_0)\delta(q^2)\ e^{-iq\cdot(x-y)}\nonumber\\
 \langle 0|\psi(x)_\beta\bar{\psi}(y)_\alpha|0\rangle &=& 
 S^{(+)}_{\beta\alpha}(x-y) = \int \frac{d^4p}{(2\pi)^3}\theta(p_0)
 \left(\mbox{$p \hspace{-0.45em}/$} + m\right)_{\beta\alpha}
 \delta(p^2-m^2)\ e^{-ip\cdot(x-y)} \nonumber\\ 
 \langle 0|\bar{\psi}(x)_\beta\psi(y)_\alpha|0\rangle &=& 
 S^{(-)}_{\beta\alpha}(x-y) = \int \frac{d^4p}{(2\pi)^3}\theta(p_0)
 \left(\mbox{$p \hspace{-0.45em}/$} - m\right)_{\beta\alpha}
 \delta(p^2-m^2)\ e^{-ip\cdot(x-y)} \nonumber\\ &\equiv& 
	\Delta^{(+)}(x-y; \mu^2).       
\label{app:A.5}\end{eqnarray}
which are the so called Wightman-functions for free-fields.
For the massive vector field $V_\mu(x)$ we have 
\begin{eqnarray}
 \langle 0|V_\mu(x)V_\nu(y)|0\rangle &=& \Delta^{(+)}_{\mu\nu}(x-y; m_V^2) = 
 \int\frac{d^4q}{(2\pi)^3}\theta(q_0)\delta\left(q^2-m_V^2\right)\ e^{-iq\cdot(x-y)}
 \left(-g_{\mu\nu}+\frac{q_\mu q_\nu}{m_V^2}\right)\ .           
\label{app:A.6}\end{eqnarray}

In the Kadyshevsky-graph theory the considered Hilbert-space is enlarged by 
admitting states containing 'quasi-particles'. The latter carry only 
4-momentum, and serve to have formally four-momentum conservation at
each vertex. The quasi-particles refer to the $\kappa$-variables in the 
Fourier transforms (\ref{app:A.4}) of the $\theta$-functions appearing in
(\ref{app:A.3}). These quasi-particle states $|\kappa_1,\ldots \rangle$ are
normalized by
\begin{equation}
 \langle \kappa'_1 \ldots | \kappa_1,\ldots\rangle = 
 \delta(\kappa'_1-\kappa_1) \ldots 
\label{app:A.7}\end{equation}
The $\theta$-functions in (\ref{app:A.3}) connect only internal points of
the graphs. In order to handle integral equations, occurring in for example
the Bethe-Salpeter- and Schwinger-Dyson-equations, one needs to consider 
amplitudes with external quasi-particles as well as internal quasi-particles.
The external quasi-particle entering a vertex is included only into the 
four-momentum conservation rule of that vertex, including both the external
and the internal quasi-particle 4-momentum.

After these preliminary remarks we now list in appendix~\ref{app:A.3}
the momentum-space rules for 
the computation of the $-M_{\kappa',\kappa}$-amplitudes, defined by
\begin{equation}
 S_{\kappa',\kappa} = 1_{\kappa',\kappa}
 -(2\pi)^4 i \delta^{4}(P_f+\kappa' n -P_i- \kappa n)\
 M_{\kappa',\kappa}.   
\label{app:A.8}\end{equation}

\section{Second Quantization Momentum Quasi-particles}
\label{app:B}
For the inclusion of the $\theta[n\cdot(x_{i-1}-x_i)]$-factors appearing in 
(\ref{app:A.3}) one may proceed as follows. Introducing $\tau'= n\cdot x',\ 
\tau= n\cdot x$ and consider the $\pi$-problem  
\begin{equation}
 \left( i\frac{\partial}{\partial\tau} + i\epsilon\right) \chi_\kappa(\tau) = 
 \kappa\ \chi_\kappa(\tau)\ .
\label{app:B.1}\end{equation}
The orthonormal-normal solutions of equation (\ref{app:B.1}) are
\begin{equation}
 \chi_\kappa(\tau) = \frac{1}{\sqrt{2\pi}}e^{-i(\kappa -i\epsilon)\tau}\ ,\ 
 -\infty < \kappa  < \infty\ .
\label{app:B.2}\end{equation}
The corresponding Green function satisfies the equation
\begin{equation}
 \left( i\frac{\partial}{\partial\tau'} + i\epsilon\right) G(\tau',\tau) = -\delta(\tau'-\tau)\ ,
\label{app:B.3}\end{equation}
which can be expressed as
\begin{eqnarray}
 G(\tau',\tau)&=& -\int_{-\infty}^\infty\ d\kappa\ \frac{\chi_\kappa(\tau')\chi_\kappa^*(\tau)}
 {\kappa + i\epsilon} = i \theta(\tau'-\tau)\ .
\label{app:B.4}\end{eqnarray}
This last expression follows from (\ref{app:B.2}) and the representation (\ref{app:A.4}).
Notice that we can also write for G the expression
\begin{eqnarray}
 G(\tau',\tau)&=& -\int_{C_R} \frac{d\kappa}{\kappa}\ \chi_\kappa(\tau')\chi_\kappa^*(\tau)\ ,
\label{app:B.5}\end{eqnarray}
where the contour $C_R$ in the complex $\kappa$-plane is $C_R=\{ -\infty < \Re\kappa < \infty,
 \Im\kappa = i\epsilon\}$.

\noindent For the second quantization formalism we introduce 
 auxiliary fields, henceforth called Kadyshevsky fields, by the operators
\begin{eqnarray}
 \chi(\tau) &=& \int\frac{d\kappa}{\kappa+i\epsilon}\
 a(\kappa)\ \chi_\kappa(\tau)\ , \nonumber\\
 \bar{\chi}(\tau) &=& \int\frac{d\kappa}{\kappa-i\epsilon}\
 a^\dagger(\kappa)\ \chi_\kappa^*(\tau)\ .
\label{app:B.6}\end{eqnarray}
In second quantization, we postulate the commutator
\footnote{ Note that one can also use anti-commutators: 
Postulating the anti-commutator
\begin{eqnarray*}
	\big\{\chi(\tau'),\bar{\chi}(\tau)\big\} =
 -i \theta(\tau'-\tau) \equiv -i \theta[n\cdot(x'-x)],
\end{eqnarray*}
the annihilation and creation operators satisfy
	$\big\{a(\kappa'),a^\dagger(\kappa)\big\} = \kappa \delta(\kappa'-\kappa)$.
	This makes the Kadyshevsky field $\chi(\tau)$ fermionic, which
	might be advantageous using path-integral techniques.
}

\begin{equation}
 \left[ \chi(\tau'), \bar{\chi}(\tau) \right] = 
 -i \theta(\tau'-\tau) \equiv -i \theta[n\cdot(x'-x)]\ ,
\label{app:B.7}\end{equation}
which follows from the canonical commutation rules for the annihilation and creation 
operators for the quasi-particles
\begin{equation}
 \left[a(\kappa'), a^\dagger(\kappa) \right] = \kappa \delta(\kappa'-\kappa)\ .
\label{app:B.8}\end{equation}
We note that with these normalizations
\begin{equation}
 |\kappa\rangle = a^\dagger(\kappa)|0\rangle\ \ , \ \
 \chi(\tau)|\kappa\rangle = \frac{1}{\sqrt{2\pi}}e^{-i(\kappa-i\epsilon)\tau}\ .
\label{app:B.9}\end{equation}

Next we introduce the following addition to the free Lagrangian density
\begin{equation}
 {\cal L}_K = i \bar{\chi}(\tau)\dot{\chi}(\tau) + i\epsilon \bar{\chi}(\tau)\chi(\tau)\ ,
\label{app:B.10}\end{equation}
where $\dot{\chi} := \partial\chi/\partial\tau$.
To the interaction Lagrangians we add a factor $\chi^\dagger\chi$, for example
for the pseudo-scalar pion-nucleon interaction
\begin{eqnarray}
 {\cal L}_{ps} = g \bar{\psi}(x)\gamma_5\psi(x)\ \phi(x) & \rightarrow &
 \bar{\cal L}_{ps} = g \left[\bar{\psi}(x)\gamma_5\psi(x)\ \phi(x)\right]\cdot
 \left\{\bar{\chi}(n\cdot x)\chi(n\cdot x)\right\}\ .
\label{app:B.11}\end{eqnarray}
This additional factor will produce in the contractions between the vertices of a graph
the factor
\begin{equation}
 \langle 0 | \chi(n\cdot x')\bar{\chi}(n\cdot x)|0\rangle = -i \theta[n\cdot(x'-x)]\ .
\label{app:B.12}\end{equation}
With these changes in the Lagrangian etc. one can formally incorporate the $\theta$-functions
appearing in (\ref{app:A.3}) in a second-quantization formalism as follows. First we write 
(\ref{app:A.3}) in the equivalent form
\begin{eqnarray}
 S_n &=& \frac{i^n}{n!} \sum_P
 \int_{-\infty}^{+\infty} \ldots \int_{-\infty}^{+\infty} 
 \theta[n\cdot(x_{\pi_1}-x_{\pi_2})] \theta[n\cdot(x_{\pi_2}-x_{\pi_3})]
 \ldots \theta[n\cdot(x_{\pi{n-1}}-x_{\pi_{n}})] 
 \cdot\nonumber\\ && \hspace{2cm} \times
 {\cal L}_I(x_{\pi_1})\ {\cal L}_{I}(x_{\pi_2}) \ldots {\cal L}_I(x_{\pi_n}) 
\cdot d^4x_1 \ldots d^4x_n\ ,            
\label{app:B.13a}\end{eqnarray}
where the sum P includes all permutation $\pi(1,2, \ldots ,n)$. Then, in the $\kappa$-space one
next defines the $S_n$-operator by 
\begin{eqnarray}
	\hspace{-3mm} \langle \kappa'| S_n| \kappa\rangle &=& \frac{i^n}{n!} \sum_P
 \int_{-\infty}^{+\infty} \ldots \int_{-\infty}^{+\infty}\ \langle \kappa'|\
 \bar{\cal L}_I(x_{\pi_1})\ \bar{\cal L}_{I}(x_{\pi_2}) \ldots \bar{\cal L}_I(x_{\pi_n}) 
 | \kappa\rangle\ \cdot d^4x_1 \ldots d^4x_n\ ,            
\label{app:B.14a}\end{eqnarray}
where the change 
 ${\cal L}_I(x) \rightarrow \bar{\cal L}_{I}(x)$ symbolizes the change in the interaction 
Lagrangians similar to that in (\ref{app:B.9}).
Taking matrix elements of the expression in (\ref{app:B.12}) generates all 
Kadyshevsky-graphs as defined by the rules in Appendix~\ref{app:B}.

\noindent The matrix elements of the full $S$-operator can now be expressed as
\begin{eqnarray}
 \langle \kappa'| S| \kappa\rangle &=& {\cal S} \exp \left\{\vphantom{\frac{A}{A}} 
 \frac{i}{\hbar}\int_{-\infty}^{+\infty} \bar{\cal L}_I(x)d^4x \right\}\ ,
\label{app:B.13b}\end{eqnarray}
where ${\cal S}$ stands for the symmetrizer
\begin{equation}
 {\cal S} \left(\vphantom{\frac{A}{A}} 
 {\cal L}_I(x_{1})\ {\cal L}_{I}(x_{2}) \ldots {\cal L}_I(x_{n})\right) = \sum_P
 {\cal L}_I(x_{\pi_1})\ {\cal L}_{I}(x_{\pi_2}) \ldots {\cal L}_I(x_{\pi_n})\ . 
\label{app:B.14b}\end{equation}

\section{Functional Integral Formalism}
\label{app:C}

\subsection{Path Integral Formalism}
\label{app:Ca}
We consider the scalar field theory. Then the Lagrangian including the Kadyshevsky fields is
\begin{equation}
 {\cal L}(x) = {\cal L}(x) + {\cal L}_K(n;x) = 
 {\cal L}_0(x) + \bar{\cal L}_I(x) + {\cal L}_K(n;x)\ , 
\label{app:C.1}\end{equation}
with  
\begin{eqnarray}
 {\cal L}_K(\chi^\dagger,\chi) &=& i\bar{\chi}(\tau)\dot{\chi}(\tau)+i\epsilon 
 \bar{\chi}(\tau)\chi(\tau)\ , \nonumber\\
 {\cal L}_0(\phi,\partial_\mu\phi) &=& \frac{1}{2}\partial_\mu\phi\partial^\mu\phi 
 -\frac{1}{2} m^2 \phi^2\ , \nonumber\\
 \bar{\cal L}_I(\phi,\chi,\bar{\chi}) &=&  - U(\phi)\cdot\bar{\chi}_n\chi_n\ , 
\label{app:C.2}\end{eqnarray}
where $\tau = n\cdot x$. 
The path integral for the generating functional reads
\begin{eqnarray}
 \langle \kappa'| Z(J,\eta_n,\eta^\dagger_n)| \kappa\rangle &=& N \int d\phi d\bar{\chi} d\chi\ 
 \exp\left\{ \frac{i}{\hbar} \int d^4x\ \left( \vphantom{\frac{A}{A}} 
 {\cal L}(x) + {\cal L}_K(n,x) + J(x)\phi(x) \right.\right. \nonumber\\
  && \left.\left. \vphantom{\frac{A}{A}} + \bar{\chi}_n(x)\eta_n(x)
 + \bar{\eta}_n(x)\chi_n(x) \right)\right\}\ .
\label{app:C.3}\end{eqnarray}

\subsection{Schwinger-Symanzik Equations}
\label{app:Cb}
In the following we omit $\langle \kappa'| \ldots | \kappa\rangle$, unless explicitly needed.
Writing (\ref{app:C.3}) as
\begin{eqnarray}
  Z(J,\eta_n,\eta^\dagger_n) &=& N \int d\phi d\bar{\chi} d\chi\ 
 \hat{Z}(\phi,\bar{\chi},\chi)\cdot \nonumber\\
 && \times\exp\left\{ \frac{i}{\hbar} \int d^4x\ \left( \vphantom{\frac{A}{A}} 
  J(x) \phi(x) + \bar{\chi}_n(x)\eta_n(x)
 + \bar{\eta}_n(x)\chi_n(x) \right) \right\}\ ,
\label{app:C.4}\end{eqnarray}
where $\hat{Z}$ in terms of the action $S[\dots]$ of the fields is given by
\begin{equation}
 \hat{Z}(\phi,\bar{\chi},\chi) = N \exp\left\{(i/\hbar)\ S\left[\phi,\bar{\chi},\chi\right]
 \vphantom{\frac{A}{A}}\right\}\ .
\label{app:C.5}\end{equation}
One finds from the field equations $\delta S[\phi,\bar{\chi},\chi]/\delta \phi(x)=0$ 
that $\hat{Z}$ satisfies the functional differential equation
\begin{eqnarray}
 i\hbar \frac{\delta \hat{Z}(\phi,\bar{\chi},\chi)}{\delta \phi(x)} &=& 
(\Box_x + m^2) \phi(x) \hat{Z}(\phi,\bar{\chi},\chi) 
-\bar{\cal L}'_I(\phi,\bar{\chi},\chi) \hat{Z}(\phi,\bar{\chi},\chi)\ ,
\label{app:C.6}\end{eqnarray}
where the prime denotes differentiation w.r.t. $\phi$.
Multiplication (\ref{app:C.6}) left and right with 
\[
 \exp\left( \frac{i}{\hbar}\int d^4x \left[J(x)\phi(x)+\bar{\eta}_n(x)\chi_n(x) +
 \eta_n(x)\bar{\chi}_n(x)\right]\right)
\]
and integrating over the fields $\phi$, etc. one arrives using partial integration at the 
Schwinger-Symanzik equation 
\begin{eqnarray}
(\Box_x + m^2) \frac{\hbar}{i}\frac{\delta Z[J,\bar{\eta},\eta]}{\delta J(x)}       
-\bar{\cal L}'_I\left(\frac{\hbar}{i}\frac{\delta}{\delta J(x)},
\frac{1}{i}\frac{\delta}{\delta \bar{\eta}(x)},
\frac{1}{i}\frac{\delta}{\delta \eta(x)}\right)\ Z[J,\bar{\eta},\eta] &=& 
 J(x)\ Z[J,\bar{\eta},\eta]\ .  
\label{app:C.7}\end{eqnarray}
Analogous equations are obtained by taking functional derivatives w.r.t. $\eta(\tau)$ and
$\bar{\eta}(\tau')$.\\

\subsection{Generation Kadyshevsky graphs, Generalized Wightman-functions}
\label{app:Cc}

\noindent The generating functional $Z_0[J]$ for the scalar lines in the Kadyshevsky-graphs 
is given by 
\begin{equation}
 Z_0[J] = N \exp\left[-\frac{i}{2\hbar}\int\ J(x) \Delta^{(+)}(x-y) J(y)\ d^4x\right],\ 
 Z_0[0]=1\ .
\label{app:C.9}\end{equation}

The generating functional $Z_K[\bar{\eta},\eta]$ for the free Kadyshevsky fields,
i.e. for the quasi-particle lines in the Kadyshevsky graphs, is given by 
\begin{equation}
Z_K[\bar{\eta},\eta] = N_K \exp\left[\vphantom{\frac{A}{A}} i\int\ \bar{\eta}(\tau)\
 \theta(\tau-\tau')\ \eta(\tau')\ d\tau d\tau'\right]\ ,
\label{app:C.8}\end{equation}
where $N_K$ is such that $Z_K[0,0] = 1$. One immediately verifies that
\begin{equation}
 \left.\frac{\partial^2 Z_K[\bar{\eta},\eta]}{\partial\bar{\eta}(\tau)\partial\eta(\tau')}
\right|_{\bar{\eta}=0,\eta=0} = i \theta(\tau-\tau')\ .
\label{app:C.10}\end{equation}

Then, the conjecture would be something like:
{\it The perturbation expansion in Kadyshevsky graphs is delivered by the functional}
\begin{eqnarray}
Z[J,\bar{\eta},\eta] &=& \bar{N}\ \exp\left[\frac{i}{\hbar} \int\ 
\bar{\cal L}_I\left(\frac{\hbar}{i}\frac{\delta}{\delta J(x)},
\frac{1}{i}\frac{\delta}{\delta \bar{\eta}_n(x)},
\frac{1}{i}\frac{\delta}{\delta \eta_n(x)}\right)\ d^4x\right]\ 
 Z_0\left[J,\bar{\eta}_n,\eta_n\right]\ ,
\label{app:C.11}\end{eqnarray}
with $\bar{N}$ such that $Z[0,0,0]=1$. Here,
\begin{equation}
 Z_0\left[J,\bar{\eta}_n,\eta_n\right] := Z_K[\bar{\eta}_n,\eta_n]\cdot Z_0[J]\ .
\label{app:C.12}\end{equation}
Then, by functional differentiation we obtain the Kadyshevky analogs of the Green
functions
\begin{eqnarray}
 {\cal W}(\tau;x_1, \ldots , x_n; \tau') &:=& 
\left.\frac{\partial^{n+2} Z[J,\bar{\eta},\eta]}
 {\partial J(x_1) \ldots \partial J(x_n)\ \partial\eta(\tau')\partial\bar{\eta}(\tau) }\right|_{J=\bar{\eta}=\eta=0}\ ,
\label{app:C.13}\end{eqnarray}
which are a kind of generalized Wightman-functions.

\subsection{Kadyshevsky Reduction Formulas}
\label{app:Cd}

The Kadyshevsky amplitudes can be retrieved from these generalized Wightman-functions as 
follows. First, we observe that 
\begin{eqnarray}
 \Delta^{(+)}(x-y) &=& \int d^3 p\ f^{(+)}_{\bf p}(x)\ f^{(+)*}_{\bf p}(y)\ , \nonumber\\
 f^{(+)}_{\bf p}(x) &=& \frac{1}{\sqrt{2\omega_{\bf p}(2\pi)^3}}\ 
 \exp -i\left(\omega_{\bf p} x^0 -{\bf p}\cdot{\bf x}\right)\ .
\label{app:C.14}\end{eqnarray}
Here, $(\Box + m^2) f^{(+)}_{\bf p}(x)=0$, the normalization is
\begin{equation}
 \int d^3x\ f^{(+)*}_{\bf p'}(x)\ i\stackrel{\leftrightarrow}{\partial}_0
 f^{(+)}_{\bf p}(x) = \delta({\bf p}'-{\bf p})\ .
\label{app:C.15}\end{equation}
From (\ref{app:C.14}) and (\ref{app:C.15}) one has
\begin{eqnarray}
-i \int d^3x\ f^{(+)}_{\bf p'}(x)\ i\stackrel{\leftrightarrow}{\partial}_0
 \Delta^{(+)}(x-y) &=& f^{(+)}_{\bf p}(y)\ , \nonumber\\                 
 i \int d^3x\ f^{(+)*}_{\bf p'}(x)\ i\stackrel{\leftrightarrow}{\partial}_0
 \Delta^{(+)}(x-y) &=& f^{(+)*}_{\bf p}(y)\ .                             
\label{app:C.16}\end{eqnarray}
Using these relations we can define the operations for the reconstruction of the 
Kadyshevsky amplitudes from the generalized Wightman-functions.
\begin{enumerate}
\item[1.] \underline{ removal external scalar line}: consider the external point $x_1$ of
${\cal W}(\tau;x_1, \ldots , x_n;\tau')$, which has the structure
\[
 {\cal W}(\tau; x_1, \ldots , x_n;\tau') = \int d^3 y_1\ \Delta^{(+)}(x_1-y_1) \ldots 
\]
Then one can replace the on-shell propagator by the external wave-function by the operation
\begin{equation}
 i \int d^3x_1\ f^{(+)*}_{\bf p_1}(x_1)\ i\stackrel{\leftrightarrow}{\partial}_0
 {\cal W}(\tau;x_1, \ldots , x_n;\tau') \Rightarrow f^{(+)*}_{\bf p_1}(y_1)\ \dots                         
\label{app:C.17}\end{equation}
\item[2.] \underline{ removal external quasi-particle line}: in this case                 
${\cal W}(\tau;x_1, \ldots , x_n;\tau')$, which has the structure
\[
 {\cal W}(\tau;x_1, \ldots , x_n;\tau') = \int d \tau_1\ [i\theta(\tau-\tau_1)] \ldots 
\]
and therefore the operation
\begin{equation}
 -i \int d\tau\ \chi_\kappa(\tau) \frac{\partial}{\partial\tau} [i\theta(\tau-\tau_1)] \ldots 
 \Rightarrow \chi_\kappa(\tau_1)\ \dots                         
\label{app:C.18}\end{equation}
\end{enumerate}
Using these results, the general 'reduction' formula, here meant is the procedure to extract 
from the generalized Wightman-functions the scattering amplitude.
A formula for this, akin to the famous LSZ-formula, can indeed be written down. We get
\begin{eqnarray}
&& {\cal M}(\kappa,p_1, \ldots , p_m; q_1, \ldots , q_n,\kappa') = 
 \nonumber\\ && \hspace{1cm} 
 (-)^{m-n} \int_{-\infty}^\infty d\tau\ \int_{-\infty}^\infty d\tau'\cdot
 \prod_{i=1}^m \int d^3x_i\ \prod_{j=1}^n \int d^3y_j\cdot\
 \chi^*_{\kappa}(\tau)\stackrel{\rightarrow}{\frac{\partial}{\partial\tau}}\cdot
 f^{(+)*}_{p_i}(x_i)\stackrel{\leftrightarrow}{\partial_0}\cdot          
 \nonumber\\ && \times 
 {\cal W}(\tau; x_1, \ldots , x_m , y_1, \ldots , y_n;\tau')\cdot         
\stackrel{\leftrightarrow}{\partial_0}\ f^{(+)}_{q_j}(y_j)\cdot
 \stackrel{\leftarrow}{\frac{\partial}{\partial\tau'}} \chi_\kappa'(\tau')\ .
\label{app:C.19}\end{eqnarray}

\subsection{The Tree-graph Functional}         
\label{app:Ce}
In this part we consider the $\phi^3$-theory in order to be more definite. 
The functional for the connected graphs $Z^c[J,\bar{\eta},\eta]$ is defined by
\begin{equation}
Z^c[J,\bar{\eta},\eta] = \frac{i}{\hbar}\ \ln Z[J,\bar{\eta},\eta]\ .                 
\label{app:C.20}\end{equation}
Then, for the connected functional the Schwinger-Symanzik equation (\ref{app:C.7}) reads
\begin{eqnarray}
(\Box_x + m^2) \frac{\delta Z^c[J,\bar{\eta},\eta]}{\delta J(x)} &=& J(x)
 + \frac{g}{2}\left\{\frac{\hbar}{i}
 \frac{\delta^2 Z^c}{\delta J(x)\cdot\delta J(x)} + 
 \left(\frac{\delta Z^c}{\delta J(x)}\right)^2\right\}\ ,
\label{app:C.21}\end{eqnarray}
which, together with the boundary condition 
$\left.\delta Z^c/\delta J\right|_{J=\bar{\eta}=\eta=0}=0$
gives by functional differentiations differential equations for the connected n-point
functions. Inspection of (\ref{app:C.21}) shows that neglecting the term 
$\delta^2 Z^c/\delta J(x)\cdot\delta J(x)$ is expected to generate the 
tree-graph structure for this theory.                    
So, for the tree-graphs the Schwinger-Symanzik  
equation reads
\begin{eqnarray}
(\Box_x + m^2) \frac{\hbar}{i}\frac{\delta Z^c_B[J,\bar{\eta},\eta]}{\delta J(x)} &=& J(x)
 + \frac{g}{2} \left(\frac{\delta Z^c_B}{\delta J(x)}\right)^2\ ,
\label{app:C.22}\end{eqnarray}

   
\section{Kadyshevsky formalism for Interactions with Derivatives}
\label{app:T}
In this section the Kadyshevsky formalism for interactions 
with derivatives is formulated. In this the proper starting point for 
the S-matrix is the formula \cite{Schweber,IZ80} ($\hbar=c=1$)
\begin{eqnarray}
 S &=& \sum_{n=0}^\infty S^{(n)} =                                
\sum_{n=0}^\infty \frac{(-i)^n}{n!} 
 \int d^4x_1\ \ldots\ \int d^4x_n\                 
T\left[{\cal H}_I(x_1)\ \ldots\ {\cal H}_I(x_n)\right]\ ,
\label{app:T.1}\end{eqnarray}
which is {\bf Lorentz-invariant and therefore frame-independent}.
Now, because of the derivatives in ${\cal L}_I$ one has
\begin{equation}
{\cal H}_I = -{\cal L}_I +\Delta {\cal H}_I\ ,       
\label{app:T.2}\end{equation}
where $\Delta {\cal H}_I$ contains in general non-covariant "contact terms" (c.t.).
The contributions from these c.t.'s to the S-matrix in (\ref{app:T.1}) cancel 
against the non-covariant terms occurring in the $S'$-matrix
\begin{eqnarray}
 S' &=& \sum_{n=0}^\infty \frac{i^n}{n!} 
 \int d^4x_1\ \ldots\ \int d^4x_n\                 
T\left[{\cal L}_I(x_1)\ \ldots\ {\cal L}_I(x_n)\right]\ ,
\label{app:T.3}\end{eqnarray}
arising from the feature that the T-product is not covariant when  
${\cal L}_I$ contains derivatives. Then, the S-matrix (\ref{app:T.1}) appears
to be covariant.\\

\subsection{The covariant frame independent T$^*$-product}
\label{app:Ta}
\noindent 1.\ Following Gross and Jackiw \cite{Gross69} we employ the covariant 
$T^*$-product such that 
\begin{eqnarray}
 S &=& \sum_{n=0}^\infty \frac{i^n}{n!} 
 \int d^4x_1\ \ldots\ \int d^4x_n\                 
 T^*\left[{\cal L}_I(x_1)\ \ldots\ {\cal L}_I(x_n)\right]\ ,
\label{app:T.4}\end{eqnarray}
Using the Kadyshevsky form of the T-products, i.e. with 
$\theta(x_i^0-x_j^0) \rightarrow \theta\left[n\cdot(x_i-x_j)\right]$,
one has \cite{Gross69}
\begin{equation}
 T^*\left[{\cal L}_I(x_1)\ \ldots {\cal L}_I(x_n)\right] = 
 T\left[{\cal L}_I(x_1)\ \ldots {\cal L}_I(x_n)\right] + 
 \tau(x_1, \ldots , x_n)\ .
\label{app:T.5}\end{equation}
In a short hand notation we write (\ref{app:T.5}) as         
\begin{equation}
 T^*(x_1, \ldots , x_n; n) = T(x_1, \ldots , x_n; n) + 
 \tau(x_1, \ldots , x_n; n)\ . 
\label{app:T.6}\end{equation}
Now, we require that $T^*$-product is frame, i.e. $n^\mu$, independent.
Then, this assures that the S-matrix is Lorentz-invariant.\\
Considering variations $\delta n^\mu$ such that $\delta n^2= n\cdot\delta n =0$, 
we have the following differential equation for $\tau(x_1, ... , x_n;n)$
\begin{equation}
 P^{\alpha\beta}\frac{\delta}{\delta n^\beta} 
 T^*(x_1, \ldots , x_n; n) = P^{\alpha\beta}\frac{\delta}{\delta n^\beta} 
 T(x_1, \ldots , x_n; n) + P^{\alpha\beta}\frac{\delta}{\delta n^\beta} 
 \tau(x_1, \ldots , x_n; n) = 0\ .
\label{app:T.7}\end{equation}
Here, we introduced the projection operator
\begin{equation}
 P^{\alpha\beta} = g^{\alpha\beta} - n^\alpha n^\beta\ ,
\label{app:T.8}\end{equation}
from which follows that $n_\alpha P^{\alpha\beta} A_\beta = 0$ for any vector $A^\mu$.\\

\noindent 2.\ Since in our applications we have to deal with $S^{(2)}$, we consider
this case in detail. We have
\begin{subequations}
\begin{eqnarray}
&& S^{(2)} = (-i)^2\frac{1}{2!}\int d^4x\ \int d^4y\ 
 T\left[{\cal L}_I(x){\cal L}_I(y)\right]\ , \\
&& T[x-y;n] = \theta[n\cdot(x-y)]\ {\cal L}_I(x){\cal L}_I(y)
         +  \theta[-n\cdot(x-y)]\ {\cal L}_I(y){\cal L}_I(x)\ ,
\label{app:T.9}\end{eqnarray}
\end{subequations}
and
\begin{eqnarray}
 P^{\alpha\beta}\frac{\delta}{\delta n^\beta} T(x-y; n) &=& 
 P^{\alpha\beta}(x-y)_\beta \delta\left[n\cdot(x-y)\right]\
 \left[{\cal L}_I(x),{\cal L}_I(y)\right]\ .
\label{app:T.10}\end{eqnarray}
Now, in general one has that
\begin{eqnarray}
 \delta[n\cdot(x-y)]\ \left[{\cal L}_I(x),{\cal L}_I(y)\right] &=&
C(n) \delta^4(x-y) +P^{\alpha\beta} S_\alpha(n) \partial_\beta \delta^4(x-y)
 \nonumber\\ && + P^{\alpha\beta}P^{\gamma\delta} 
Q_{\alpha\gamma}(n)\partial_\beta\partial_\delta \delta^4(x-y) \dots\ ,
\label{app:T.11}\end{eqnarray}
where the $\ldots$ stand for terms with higher order derivatives of the 
$\delta$-function. The terms on the r.h.s. with the derivatives are known in
the literature as the '{\it Schwinger terms}'. 
We will refer to $S_\alpha$ and $Q_{\alpha\gamma}$ as the 'dipole' and
'quadrupole' Schwinger term respectively. In the following discussion
we ignore possible higher-order derivatives on the r.h.s. of (\ref{app:T.11}).\\
 
\noindent (i)\ IF $S_\alpha(n)=0$ etc., i.e. no-derivatives terms on the r.h.s.: 
\[                 
 P^{\alpha\beta}\frac{\delta}{\delta n^\beta} T(x-y; n) = 0\ \Rightarrow 
 \tau(x-y;n) = 0\ .
\]
\noindent (ii)\ IF $S_\alpha(n) \neq 0, Q_{\alpha\gamma}(n) = 0$, $\tau(x-y;n)$ 
satisfies the equation           
\begin{eqnarray}
 P^{\alpha\beta}\frac{\delta}{\delta n^\beta} \tau(x-y; n) &=&               
 P^{\alpha\beta}\ S_\beta(n) \delta^4(x-y)\ ,                                
\label{app:T.12}\end{eqnarray}
with the solution \cite{Gross69}
\begin{eqnarray}
 \tau(x-y;n) &=& \int^n dn_\beta'\ S^\beta(n')\ \delta^4(x-y) + \tau_0(x-y)\ .           
\label{app:T.13}\end{eqnarray}
In our applications it will appear that in this case $S^\alpha$ is such that the 
solution can be written as
\[
 \tau(x-y;n) = (S\cdot n) \delta^4(x-y) + \tau_0(x-y)\ .
\]
\noindent (iii)\ IF $S_\alpha(n), Q_{\alpha\gamma}(n) \neq 0$, $\tau(x-y;n)$ 
satisfies the equation           
\begin{eqnarray}
 P^{\alpha\beta}\frac{\delta}{\delta n^\beta} \tau(x-y; n) &=&               
 P^{\alpha\beta}\ S_\beta(n) \delta^4(x-y) \nonumber\\                       
 && + P^{\alpha\beta}P^{\gamma\delta}\left(Q_{\beta\gamma}+Q_{\gamma\beta}\right)
 \partial_\delta \delta^4(x-y)\ ,
\label{app:T.14a}\end{eqnarray}
with the solution 
\begin{eqnarray}
 \tau(x-y;n) &=& \int^n dn_\beta' \left\{ S^\beta(n') + 
 P^{\gamma\delta}(n')\left(Q_{\beta\gamma}+Q_{\gamma\beta}\right)(n')\partial_\delta
 \right\}\cdot \nonumber\\ &&
 \times\delta^4(x-y) + \tau_0(x-y)\ .            
\label{app:T.14b}\end{eqnarray}

\noindent {\it From this emerges the following scheme:\\
 
 a) With $T, {\cal L}_I\ \Rightarrow$ Covariant- + non-covariant (N.C.)- terms.\\
 
 b) With $T^*, {\cal L}_I\ \Rightarrow$ Covariant-  + N.C.- + $\tau$-terms 
 $\Rightarrow$ Covariant terms.
}\\

\subsection{The covariant frame independent Kadyshevsky formalism}
\label{app:Tb}
Next, we consider the adaption of the Kadyshevsky formalism 
necessary to cope with derivative interactions. From the analysis given above, 
it is clear that scheme b) is the proper one in order to produce covariant
S-matrix elements. Therefore, in order that the Kadyshevsky formalism yields
the same S-matrix as the Feynman formalism, we adapt the former.            
Investigating the $n^\mu$-dependence we consider
\begin{equation}
P^{\alpha\beta}\frac{\delta}{\delta n^\beta} \theta[n\cdot(x-y)] 
 {\cal L}_I(x){\cal L}_I(y) =          
P^{\alpha\beta} (x-y)_\beta \delta[n\cdot(x-y)] {\cal L}_I(x){\cal L}_I(y)
\label{app:T.15}\end{equation}
Next we make the observation/conjecture that 
\begin{subequations}
\begin{eqnarray}
	&& \delta[n\cdot(x-y)] {\cal L}_I(x){\cal L}_I(y) = 
N\left[{\cal L}_I(x){\cal L}_I(y)\right]|_0 + \frac{1}{2} C(n)\delta({\bf x}-{\bf y})
+ \frac{1}{2} S^i(n)\partial_i\delta({\bf x}-{\bf y}), \\
	&& \delta[-n\cdot(x-y)] {\cal L}_I(y){\cal L}_I(x) = 
N\left[{\cal L}_I(x){\cal L}_I(y)\right]|_0 - \frac{1}{2} C(n)\delta({\bf x}-{\bf y})
- \frac{1}{2} S^i(n)\partial_i\delta({\bf x}-{\bf y}). 
\label{app:T.16}\end{eqnarray}
\end{subequations}
Here,$N[...]$ denotes the so called normal-ordered product.
We illustrate this for the following elementary case:
\begin{eqnarray*}
	&& \phi(x) \dot{\phi}(y) = N[\phi(x)\dot{\phi}(y)] -\partial_0 
 \Delta^{(+)}(x-y;m^2)\ , \\
	&& \dot{\phi}(y) \phi(x) = N[\phi(x)\dot{\phi}(y)] +\partial_0 
 \Delta^{(+)}(x-y;m^2)\ ,
\end{eqnarray*}
which gives $C(n)/2 = -\delta({\bf x-y})$ and $S_m(n)=0$. 
Here, see \cite{Schweber} section 7c, 
\begin{eqnarray*}
 \Delta^{(+)}(x-y;m^2) &=& \frac{1}{2}\left[
 \Delta(x-y;m^2) -i\Delta^{(1)}(x-y;m^2)\right]\ \ ,\ \ {\rm with} \nonumber\\
 \partial_0\Delta(x-y;m^2)\left.\right|_0 &=& -\delta({\bf x-y})\ \ ,\ \ 
 \partial_0\Delta^{(1)}(x-y;m^2)\left.\right|_0 = 0\ .                         
\end{eqnarray*}

\noindent Next, consider the important case 
\begin{eqnarray}
 && \partial^\mu\phi(x) \partial^\nu\phi(y) = 
  N\left[\partial^\mu\phi(x) \partial^\nu\phi(y)\right] + 
 \partial^\mu\partial^\nu \Delta^{(+)}(x-y;m^2)\ ,
\label{app:T.17}\end{eqnarray}
which leads to 
\begin{subequations}\label{app:T.18}
\begin{eqnarray}
&& \delta[n\cdot(x-y)]\partial^\mu\phi(x) \partial^\nu\phi(y) = 
  N\left[\partial^\mu\phi(x) \partial^\nu\phi(y)\right]|_0
 -\left(\delta^\mu_m \delta^\nu_0+ \delta^\mu_0\delta^\nu_m\right)\
 \partial^m_x\delta({\bf x}-{\bf y})\ , \\
&& \delta[-n\cdot(x-y)]\partial^\nu\phi(y) \partial^\mu\phi(x) = 
  N\left[\partial^\mu\phi(x) \partial^\nu\phi(y)\right]|_0
 +\left(\delta^\mu_m \delta^\nu_0+ \delta^\mu_0\delta^\nu_m\right)\
 \partial^m_x\delta({\bf x}-{\bf y}), 
\end{eqnarray}\end{subequations}
which gives 
\begin{equation}
 C(n)=0\ \ ,\ \ \frac{1}{2} S_m(0) \Leftarrow 
 +\left(\delta^\mu_m \delta^\nu_0+ \delta^\mu_0\delta^\nu_m\right)\ .
\label{app:T.19}\end{equation}
{\it We note that the 'conjecture' (\ref{app:T.15}) is consistent with 
 (\ref{app:T.11}) }.\\

\noindent After these preparations, the adaption of the Kadyshevsky formalism
runs as follows:\\

\noindent a) We present the S-matrix formula (\ref{app:A.3}) in the form 
\begin{eqnarray}
 S^{(n)} &=& i^n \int d^4x_n \ldots \int d^4x_1\             
 R\left[{\cal L}_I(x_n)\ {\cal L}_{I}(x_{n-1}) \ldots {\cal L}_I(x_1)\right]\ ,
\label{app:T.20}\end{eqnarray}
where 
\begin{eqnarray}
 R\left[{\cal L}_I(x_n)\ {\cal L}_{I}(x_{n-1}) \ldots {\cal L}_I(x_1)\right]
 &=& \theta[n\cdot(x_n-x_{n-1})] \theta[n\cdot(x_{n-1}-x_{n-2})]
 \ldots \theta[n\cdot(x_{2}-x_{1})]\cdot\nonumber\\ && \times
 \left[{\cal L}_I(x_n)\ {\cal L}_{I}(x_{n-1}) \ldots {\cal L}_I(x_1)\right]\ .
\label{app:T.21}\end{eqnarray}
Similarly to the case for the T-product, we now introduce the covariant $R^*$-product,
which,restricting ourselves to the second order case, is related to the R-product by 
\begin{equation}
 R^*(x-y;n) = R(x-y;n) + \rho(x-y;n)\ , 
\label{app:T.22}\end{equation}
similarly to the definition (\ref{app:T.6}). Requiring now that the $R^*$-product is
frame-independent one obtains the equation that 
\begin{equation}
 P^{\alpha\beta}\frac{\delta}{\delta n^\beta} 
 R^*(x_1, \ldots , x_n; n) = P^{\alpha\beta}\frac{\delta}{\delta n^\beta} 
 R(x_1, \ldots , x_n; n) +  P^{\alpha\beta}\frac{\delta}{\delta n^\beta} 
 \rho(x_1, \ldots , x_n; n) = 0\ .
\label{app:T.23}\end{equation}
which gives for the two-point functions 
\begin{eqnarray}
 P^{\alpha\beta}\frac{\delta}{\delta n^\beta} \rho(x-y; n) &=&    
 \frac{1}{2}\left[ P^{\alpha\beta} S_\beta(n) 
 + P^{\alpha\beta} P^{\gamma\delta}\left(Q_{\beta\gamma}+Q_{\gamma\beta}\right)
 \partial_\delta\right]\ \delta^4(x-y)\ . 
\label{app:T.24}\end{eqnarray}
From this it follows that $\rho(x-y;n)= \tau(x-y;n)/2$.\\

\noindent {\it 
Then, similarly as in the Feynman formalism, the introduction of the $R^*$-product
in the Kadyshevsky formalism yields a covariant and frame independent S-matrix, 
and $S(Kadyshevky)=S(Feynman)$ for on-shell initial and final states.}

 \begin{figure} \begin{center}
\resizebox{8.25cm}{!}
 {\includegraphics[width=14cm,height=5cm]{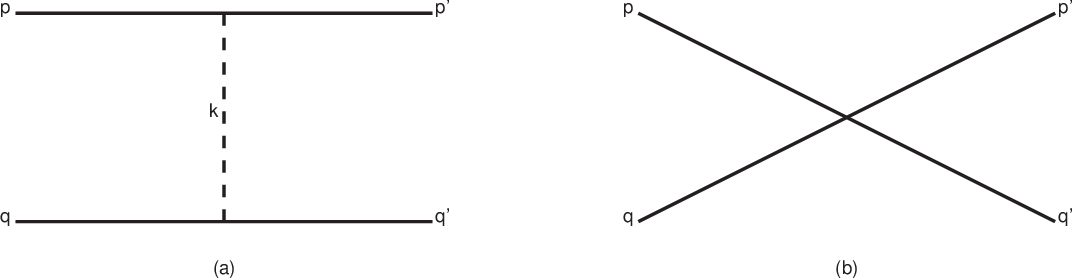}}
\caption{Panel (a): NN OPE-exchange. Panel (b): Seagull grtaph contribution.}
 \label{fig.obe-seagull}
 \end{center}\end{figure}
\section{Examples Interactions with Derivatives}
\label{app:EX}

\noindent 1.\ \underline{\bf Pseudo-vector Pion-Nucleon Interaction}: 
The interaction Lagrangian density for neutral mesons reads
\begin{equation}
 {\cal L}_{pv} = \frac{f}{m_\pi}\ \bar{\psi}(x) \gamma_5 \gamma_\mu \psi(x) 
 \partial^\mu\phi(x)\ ,
\label{app:EX.1a}\end{equation}
where the fields are in the 'interaction representation', for which we take the 
'in'-fields. This means they satisfy the free field commutation relations.
The free Hamiltonian density is
\begin{eqnarray}
	{\cal H}_0 &=& \bar{\psi}(x)\bigl(\bm{\gamma}\cdot{\bf p}+M\bigr)\ \psi(x)
	+\frac{1}{2}\bigl(\pi^2({\bf x})+\bm{\nabla}\phi\cdot\bm{\nabla}\phi(\bf x)
	+m_\pi^2 \phi^2({\bf x})\bigr)
\label{app:EX.1b}\end{eqnarray}
The canonical momentum for the pion is
\begin{eqnarray}
	\pi &=& \frac{\partial {\cal L}}{\partial\dot{\phi}} = \dot{\phi}-\frac{f}{m_\pi}
	\psi^\dagger \gamma_5 \psi
\label{app:EX.1c}\end{eqnarray}
which, after eliminating $\dot{\phi}$, implies that the interaction Hamiltonian density becomes
\begin{eqnarray}
	{\cal H}_I(x) &=& -{\cal L}_{pv} -2M\left(\frac{f}{m_\pi}\right)^2 \bar{\psi}(x) \psi(x) \phi^2(x)
+\frac{1}{2}\left(\frac{f}{m_\pi}\right)^2 \left(\psi^\dagger(x)\gamma_5 \psi(x)\right)^2,
\label{app:EX.1d}\end{eqnarray}
where the second term arises from the elimination of the virtual nucleon-antinucleon 
pair transition. The third term is the "contact" interaction which is characteristic
for a Lagrangian with a derivative coupling \cite{Schweber}.
This non-covariant contact term cancels the non-covariant term in second order S-matrix 
for nucleon-nucleon scattering, see Fig.~\ref{fig.obe-seagull}.

\noindent The equal-time anti-commutation (ETAC) and commutation (ETC) relation are, respectively
\begin{subequations}
\begin{eqnarray}
 \left\{\psi_a(x), \psi^*_b(y)\right\}|_0 &=& \delta_{a,b}\delta({\bf x-y})\ , \\
 \left[\phi(x), \dot{\phi}(y)\right]|_0 &=& \delta({\bf x-y})\ ,    
\label{app:EX.2}\end{eqnarray}
\end{subequations}
and the other ETC's etc.are zero.
The ETC relation for the interaction Lagrangian is
\begin{equation}
\left[{\cal L}_{pv}(x), {\cal L}_{pv}(y)\right]|_0 = \frac{f^2}{m_\pi^2}
\left(\Gamma_\mu\right)_{ab} \left(\Gamma_\nu\right)_{cd}
 \left[\psi_a^\dagger\psi_b(x)\partial^\mu\phi(x) , 
 \psi_c^\dagger\psi_d(y)\partial^\nu\phi(y)\right]|_0 
\label{app:EX.3}\end{equation}
Using the commutator
\begin{equation}
 \left[ABF,CDG\right] = ABCD\ [F,G] + [AB,CD]\ GF\ ,                                  
\label{app:EX.4}\end{equation}
where is supposed that F and G commute with $A,B,C,D$. We identify $A-D$ with
the nucleon- and (F,G) with the pion-operators. Furthermore, we have that
\begin{equation}
 \left[AB,CD\right] = A \left\{B,C\right\} D - C \left\{A,D\right\} B
\label{app:EX.5}\end{equation}
which holds when $\{A,C\} = \{B,D\}=0$. Then, we easily derive that
\begin{subequations}
\begin{eqnarray}
&& \left[ \psi_a^\dagger \psi_b(x) , \psi_c^\dagger \psi_d(y)\right]|_0 =
 \left(\psi_a^\dagger \delta_{bc} \psi_d - \psi_c^\dagger \delta_{ad} \psi_b\right)\
 \delta({\bf x-y})\ , \\
&& \left[\partial^\mu\phi(x), \partial^\nu\phi(y)\right]|_0 =
 i \left(\delta^\mu_m\delta^\nu_0 - \delta^\mu_0 \delta^\nu_n\right)
 \delta({\bf x-y})\ .       
\label{app:EX.6}\end{eqnarray}
\end{subequations}
Using these results we get that
\begin{eqnarray*}
&& \left[\psi_a^\dagger\psi_b(x)\partial^\mu\phi(x) , 
 \psi_c^\dagger\psi_d(y)\partial^\nu\phi(y)\right]|_0 = \nonumber\\
&& i \left(\psi^\dagger_a\psi_b\cdot\psi_c^\dagger \psi_d\right)|_0
 \left(\delta^\mu_m\delta^\nu_0 - \delta^\mu_0 \delta^\nu_n\right)
 \delta({\bf x-y}) + \nonumber\\ &&
 \left(\psi_a^\dagger\delta_{bc}\psi_d-\psi_c^\dagger\delta_{ad}\psi_b\right)\
 \partial^\nu\phi(y)\partial^\mu\phi(x) \delta({\bf x-y})\ .                
\end{eqnarray*}
The second term in this commutator gives a term proportional to
\begin{eqnarray*} 
&&  \psi^\dagger\left(\Gamma_\mu\Gamma_\nu - \Gamma_\nu \Gamma_\mu\right)\psi 
\Rightarrow \bar{\psi}\left(\gamma_\mu\gamma_0\gamma_\nu -\gamma_\nu\gamma_0\gamma_\mu
\vphantom{\frac{A}{A}}\right)\psi\ ,
\end{eqnarray*}
which on inspection vanishes for either $\mu=0$ or $\nu=0$. For $\mu=m,\nu=n$ it becomes
\begin{eqnarray*}
&& \psi^\dagger\left(\gamma_n\gamma_m-\gamma_m\gamma_n\right)\psi\
\partial^m\phi(x)\partial^n\phi(y)|_0\ \delta^4(x-y) \Rightarrow 0\ .
\end{eqnarray*}
Finally, we obtain for the commutator of the interaction Lagrangian
\begin{eqnarray}
&& \left[{\cal L}_{pv}(x), {\cal L}_{pv}(y)\right]|_0 = \frac{f^2}{m_\pi^2}
\left(\psi^\dagger\Gamma_\mu\psi\right)\left(\psi^\dagger\Gamma_\nu\psi\right)
 \left[\partial^\mu\phi(x),\partial^\nu\phi(y)\right]|_0\ \nonumber\\
&& i \frac{f^2}{m_\pi^2}\left[\psi^\dagger\Gamma_m\psi\cdot\psi^\dagger\Gamma_0\psi\
+\psi^\dagger\Gamma_0\psi\cdot\psi^\dagger\Gamma_m\psi\partial^n_y\right]|_0
\partial^m_x\delta({\bf x-y})\ .              
\label{app:EX.7}\end{eqnarray}
Multiplying this result with $\delta(x^0-y^0)$ we infer that the r.h.s contains       
a Schwinger term, which leads to
\begin{eqnarray}
 S_\mu(n) &=& i \frac{f^2}{m_\pi^2}
\left[\psi^\dagger\Gamma_\mu\psi\cdot\psi^\dagger\Gamma_0\psi\
+\psi^\dagger\Gamma_0\psi\cdot\psi^\dagger\Gamma_\mu\psi\right]|_0 \nonumber\\
 &=& i \frac{f^2}{m_\pi^2}
\left[\psi^\dagger\Gamma_\mu\psi\cdot\psi^\dagger(\Gamma\cdot n)\psi\
+\psi^\dagger(\Gamma\cdot n)\psi\cdot\psi^\dagger\Gamma_\mu\psi\right]|_0\ .            
\label{app:EX.8}\end{eqnarray}
Then, taking $n^\mu=(1,{\bf 0})$ we get
\begin{eqnarray}
 \tau(x-y;n) &=& (S\cdot n) \delta^4(x-y) =  
 2i \frac{f^2}{m_\pi^2}\left(\psi^\dagger \gamma_5 \psi\right)^2 \delta^4(x-y)\ .
\label{app:EX.9}\end{eqnarray}
This is indeed the right $\tau(x-y;n)$ to cancel the non-covariant term produced
by the T-product in second order.\\

\noindent 2.\ \underline{\bf Vector-exchange in Pion-Nucleon}: 
The interaction Lagrangian we take in this case is
\begin{equation}
 {\cal L}_I(x) = g \bar{\psi}\gamma_\mu\psi\ V^\mu + f 
 \phi^\dagger\stackrel{\leftrightarrow}{\partial}_\mu \phi\ V^\mu 
 \equiv {\cal L}_I^{(1)}(x) + {\cal L}_I^{(2)}(x)
\label{app:EX.11}\end{equation}
For the vector field we have the ETC
\begin{equation}
 \left[V^\mu(x), V^\nu(y) \right] = i \left(-g^{\mu\nu}
 -\frac{\partial^\mu\partial^\nu}{m_V^2}\right) \Delta(x-y;m_V^2)
\label{app:EX.12}\end{equation}
giving
\begin{equation}
 \left[V^\mu(x), V^\nu(y) \right]|_0 = \frac{i}{m_V^2} \left(
 \delta^\mu_m\delta^\nu_0+\delta^\mu_0\delta^\nu_0\right)
 \partial^m\delta({\bf x-y})\ .
\label{app:EX.13}\end{equation}
This gives for the relevant Lagrangian commutator for $\pi N$ 
\begin{eqnarray}
&& \left[{\cal L}_I^{(1)}(x),{\cal L}_I^{(2)}(y)\right]|_0 = 
 i\frac{fg}{m_V^2}\left[\bar{\psi}\gamma_m\psi\cdot
 \phi^\dagger\stackrel{\leftrightarrow}{\partial}_0\phi  
 +\bar{\psi}\gamma_0\psi\cdot
 \phi^\dagger\stackrel{\leftrightarrow}{\partial}_m\phi\right]
 \partial^m\delta({\bf x-y})\ .
\label{app:EX.14}\end{eqnarray}
This leads for $n^\mu=(1,{\bf 0})$ to the Schwinger term
\begin{eqnarray}
&& S_\alpha = i\frac{fg}{m_V^2}\left[\bar{\psi}\gamma_\alpha\psi\cdot
 \phi^\dagger\stackrel{\leftrightarrow}{\partial}_0\phi  
 +\bar{\psi}\gamma_0\psi\cdot
 \phi^\dagger\stackrel{\leftrightarrow}{\partial}_\alpha\phi\right]\ .
\label{app:EX.15}\end{eqnarray}
Then, for $n^\mu=(1,{\bf 0})$ we find
\begin{equation}
\tau(x-y;n) = (S\cdot n) = 
 2i\frac{fg}{m_V^2}\left[\bar{\psi}(\gamma\cdot n)\psi\right]\
 \left(\phi^\dagger(\stackrel{\leftrightarrow}{\partial}\cdot n)\phi\right)\
 \delta^4(x-y)\ .
\label{app:EX.16}\end{equation}
This is indeed the correct term to cancel the non-covariant piece from the
second-order.\\

\noindent Now, it is obvious that for a vector-baryon-baryon interaction of the form
\begin{equation}
 {\cal L}_{BBV}(x) = \left[ F_1^V \bar{\psi}\gamma_\mu\psi  
+ i F_2^V \bar{\psi}\stackrel{\leftrightarrow}{\partial}_\mu \psi\right] V^\mu\ ,
\label{app:EX.17}\end{equation}
in (\ref{app:EX.16}) for the corresponding result one must make the substitution
\begin{equation}
  g\ \gamma\cdot n \rightarrow \left[ F_1^V \gamma\cdot n + 
  i F_2^V \stackrel{\leftrightarrow}{\partial}\cdot n\right]\ .
\label{app:EX.18}\end{equation}

\noindent 3.\ \underline{\bf Gauge-invariant $\Delta_{33}N\pi$ interaction}: 
 The gauge-invariant $\Delta_{33}N\pi$ interaction Lagrangian reads
\begin{eqnarray}
 {\cal L}_I &=& g \epsilon^{\mu\nu\alpha\beta}\left[
\left(\partial_\mu\bar{\psi}_\nu\right)\Gamma_\alpha \psi\cdot\partial_\beta\phi
 + \bar{\psi}\Gamma_\alpha\left(\partial_\mu\psi_\nu\right)\cdot\partial_\beta\phi
\right] \nonumber\\ &\equiv& {\cal L}^{(1)}_I + {\cal L}^{(2)}_I\ ,
\label{app:EX.21}\end{eqnarray}
with, $\Gamma_\alpha=\gamma_5\gamma_\alpha$ and, of course, 
\begin{equation}
{\cal L}^{(2)\dagger}_I = {\cal L}^{(1)}_I\ .
\label{app:EX.22}\end{equation}
The equal-time (anti)commutation relations for the nucleon and pion fields are
\begin{subequations}
\begin{eqnarray}
 \left\{\psi_a(x), \psi_b^\dagger(y)\right\}|_0 &=& 
 \delta_{a b}\delta({\bf x-y})\ , \nonumber\\ 
\left[\phi(x), \dot{\phi}(y)\right]|_0 &=& i\delta({\bf x-y})\ .
\label{app:EX.23a}\end{eqnarray}
\end{subequations}
For the (free) Rarita-Schwinger field \cite{RS41} field one has \cite{Car71}
\begin{subequations}
\begin{eqnarray}
 \left\{\psi_a^\nu(x), \bar{\psi}_b^\lambda(y)\right\} &=& (-i)
\Lambda_{\nu\lambda}(-i\partial)\left(i\gamma\cdot\partial+M\right)\Delta(x-y;M)\ ,
 \nonumber\\ 
\Lambda_{\nu\lambda}(-i\partial) &=& g_{\nu\lambda}-\frac{1}{3}\gamma_\nu\gamma_\lambda
+\frac{2}{3M^2}\partial_\nu\partial_\lambda 
+\frac{i}{3M}\left(\gamma_\nu\partial_\lambda-\gamma_\lambda\partial_\nu\right)\ .
\label{app:EX.23b}\end{eqnarray}
\end{subequations}
\noindent In the following, we use the commutator formulas
\begin{eqnarray}
\left[A B F, C D G\right] &=& ABCD \left[F,G\right] + \left[AB,CD\right]\ GF\ ,
\nonumber\\ 
\left[AB,CD\right] &=& A\left\{B,C\right\} D - C\left\{A,D\right\} B\ ,
\label{app:EX.24}\end{eqnarray}
for the case where (i) F an G commute with the set (A,B,C,D), and (ii)
$\{A,C\}=\{B,D\}=0$. Below we identify for example 
$A=\partial_\mu\psi_\nu^\dagger(x), B= \psi(x)$, and $C=\psi^\dagger(y), 
D=\partial_\rho\psi_\lambda(y)$, and  
$F=\partial_\beta\phi(x), G=\partial_\delta\phi(y)$.\\

\noindent For $\pi N$-scattering in second order in g, we get only contributions from 
the cross-term commutators $[{\cal L}^{(2)}_I(x),{\cal L}^{(1)}_I(y)]$. 
and $[{\cal L}^{(1)}_I(x),{\cal L}^{(2)}_I(y)]$, henceforth referred to as the 21- 
and the 12-commutator respectively. The 21-commutator gives 
\begin{eqnarray}
&& \left[{\cal L}_I^{(2)}(x), {\cal L}_I^{(1)}(y)\right] =            
 g^2\epsilon^{\mu\nu\alpha\beta}\epsilon^{\rho\lambda\gamma\delta}\
\left[\bar{\psi}\Gamma_\alpha\left(\partial_\mu\psi_\nu\right)\cdot\partial_\beta\phi(x),
\left(\partial_\rho\bar{\psi}_\lambda\right)\Gamma_\gamma\psi\cdot\partial_\delta\phi(y)
\right] \Rightarrow 
\nonumber\\ && 
 g^2\epsilon^{\mu\nu\alpha\beta}\epsilon^{\rho\lambda\gamma\delta}
 \left(\partial_\beta\phi(x)\ \partial_\delta\phi(y)\right)\  
 \left(\Gamma_\alpha\right)_{ab} \left(\Gamma_\gamma\right)_{cd}\cdot
\bar{\psi}_a(x)\left\{\partial_\mu\psi_{\nu,b}(x),
\partial_\rho\bar{\psi}_{\lambda,c}(y)\right\}\psi_d(y) =
\nonumber\\ && 
 +i g^2\epsilon^{\mu\nu\alpha\beta}\epsilon^{\rho\lambda\gamma\delta}
 \left(\partial_\beta\phi(x)\ \partial_\delta\phi(y)\right)\  
 \left(\Gamma_\alpha\right)_{ab} \left(\Gamma_\gamma\right)_{cd}
\cdot\nonumber\\ && 
\times\bar{\psi}(x)\left[\partial^x_\mu\partial_\rho^x\Lambda_{\nu\lambda}(-\partial)
\left(i\mbox{$\partial \hspace{-0.45em}/$}+M_\Delta\right) \Delta(x-y;M_\Delta^2)
\right]_{bc}\psi_d(y) =
\nonumber\\ && 
 +i g^2\epsilon^{\mu\nu\alpha\beta}\epsilon^{\rho\lambda\gamma\delta}
 \left(\partial_\beta\phi(x)\ \partial_\delta\phi(y)\right)\  
 \left(\Gamma_\alpha\right)_{ab} \left(\Gamma_\gamma\right)_{cd}
\cdot\nonumber\\ && 
\times\bar{\psi}(x)\left[\left(g_{\nu\lambda}-\frac{1}{3}\gamma_\nu\gamma_\lambda\right)
\partial^x_\mu\partial_\rho^x
\left(i\mbox{$\partial \hspace{-0.45em}/$}+M_\Delta\right) \Delta(x-y;M_\Delta^2)
\right]_{bc}\psi_d(y)\ . 
\label{app:EX.25}\end{eqnarray}
Insertion of the $\Gamma_{\alpha,\gamma}$ gives 
\begin{eqnarray}
&& \left[{\cal L}_I^{(2)}(x), {\cal L}_I^{(1)}(y)\right] =            
-\frac{i}{3} g^2\epsilon^{\mu\nu\alpha\beta}\epsilon^{\rho\lambda\gamma\delta}
 \left(\partial_\beta\phi(x)\ \partial_\delta\phi(y)\right)  
\cdot\nonumber\\ && 
\times\bar{\psi}(x)\gamma_\alpha 
\left[\left(g_{\lambda\nu}+\gamma_\lambda\gamma_\nu\right)
\left(i\mbox{$\partial \hspace{-0.45em}/$}-M_\Delta\right) 
\partial^x_\mu\partial_\rho^x \Delta(x-y;M_\Delta^2)
\right]\gamma_\gamma \psi(y)\ .
\label{app:EX.26}\end{eqnarray}
Taking now equal times, i.e. $x^0=y^0$, then only the time-derivatives operating
on the invariant $\Delta(x-y;M_\Delta^2)$-function survive. There will be four  
contributions: (i) $i\gamma^0\partial_0$ from the Dirac-operator, (ii)
$\delta^0_\mu\partial_0\partial_{\rho=r}$, (iii)
$\delta^0_\rho\partial_0\partial_{\mu=m}$, and (iv) for $\mu=\rho=0$ giving
$i\gamma^0\partial_0\cdot\partial^2_0= i\gamma^0\partial_0\cdot            
\sum_{k=1,3}\partial_k\partial^k-M_\Delta^2)~i\gamma^0\partial_0\cdot\sum_{k=1,3}
\partial_k\partial^k$. In the last step we skipped the $M_\Delta^2$-term since
it will not give a Schwinger term and therefore can be ignored.

To write the spatial derivatives
in a covariant form we use the $n^\mu$-vector, and write $\partial_m= P_{\mu\kappa}
\partial^\kappa$. We find
\begin{eqnarray}
&& \delta\left[n.(x-y)\right] \left[{\cal L}_I^{(2)}(x), {\cal L}_I^{(1)}(y)\right] =
-\frac{i}{3} g^2\epsilon^{\mu\nu\alpha\beta}\epsilon^{\rho\lambda\gamma\delta}
\cdot\nonumber\\ && \times
 \left(\partial_\beta\phi(x)\ \partial_\delta\phi(y)\right)\cdot\left[
 \vphantom{\frac{A}{A}}
\bar{\psi}(x)\gamma_\alpha \left(g_{\lambda\nu}+\gamma_\lambda\gamma_\nu\right)
\right.\nonumber\\ && 
\times\left\{ \vphantom{\frac{A}{A}}
-i(\gamma\cdot n)\ \left(P_{\mu\kappa}P_{\rho\sigma}
 -n_\mu n_\rho P_{\kappa\sigma}\right) \partial^\kappa\partial^\sigma \right.
\nonumber\\ && \left.
-\left(i\gamma^\tau P_{\tau\omega}\partial^\omega-M_\Delta\right)
 \left(n_\mu P_{\rho\sigma}+n_\rho P_{\mu\sigma}\right)\partial^\sigma 
 \vphantom{\frac{A}{A}} \right\} \delta^4(x-y)
\cdot\nonumber\\ && 
\times \gamma_\gamma \psi_(y) \left.\vphantom{\frac{A}{A}}\right]\ .
\label{app:EX.27}\end{eqnarray}
In order to exhibit the dipole- and quadrupole-type Schwinger terms, 
 i.e. terms with respectively one and two derivatives on the $\delta^4(x-y)$-function,
we write the above expression as
\begin{eqnarray}
&& \delta\left[n.(x-y)\right] \left[{\cal L}_I^{(2)}(x), {\cal L}_I^{(1)}(y)\right] =
-\frac{i}{3} g^2\epsilon^{\mu\nu\alpha\beta}\epsilon^{\rho\lambda\gamma\delta}
\cdot\nonumber\\ && \times
\left(\partial_\beta\phi(x)\ \partial_\delta\phi(y)\right)\cdot \left[
\vphantom{\frac{A}{A}}\right.
\bar{\psi}(x)\gamma_\alpha \left(g_{\lambda\nu}+\gamma_\lambda\gamma_\nu\right)
\cdot\nonumber\\ && 
\times\left\{ \vphantom{\frac{A}{A}}
M_\Delta \left(n_\mu P_{\rho\sigma}+
 n_\rho P_{\mu\sigma}\right)\partial^\sigma\delta^4(x-y) \right.
\nonumber\\ && \left.
-i(\gamma\cdot n)\ \left(P_{\mu\kappa}P_{\rho\sigma}
 -n_\mu n_\rho P_{\kappa\sigma}\right) \partial^\kappa\partial^\sigma\delta^4(x-y)
\right.\nonumber\\ && \left.  -i\gamma^\tau P_{\tau\omega}
 \left(n_\mu P_{\rho\sigma}+n_\rho P_{\mu\sigma}\right)
\partial^\omega\partial^\sigma\delta^4(x-y) 
\vphantom{\frac{A}{A}} \right\}
 \times \gamma_\gamma \psi_(y)
\left.\vphantom{\frac{A}{A}}\right]\ .
\label{app:EX.28}\end{eqnarray}
We recall that from (\ref{app:T.9}) 
\begin{subequations}
\begin{eqnarray*}
&& S^{(2)} = \frac{(+i)^2}{2!}\int d^4x\ \int d^4y\ 
 T\left[{\cal L}_I(x){\cal L}_I(y)\right]\ , \\
&& T[x-y;n] = \theta[n\cdot(x-y)]\ {\cal L}_I(x){\cal L}_I(y)
         +  \theta[-n\cdot(x-y)]\ {\cal L}_I(y){\cal L}_I(x)\ ,
\end{eqnarray*}
\end{subequations}
and (\ref{app:T.10})
\begin{eqnarray*}
 P^{\alpha\beta}\frac{\delta}{\delta n^\beta} T(x-y; n) &=& 
 P^{\alpha\beta}(x-y)_\beta \delta\left[n\cdot(x-y)\right]\
 \left[{\cal L}_I(x),{\cal L}_I(y)\right]\ .
\end{eqnarray*}
that 
\begin{eqnarray}
 P_{\omega\kappa}\frac{\delta}{\delta n_\kappa}S^{(2)} &\Leftarrow& 
-\frac{1}{2!}\int d^4x\ \int d^4y\ P_{\omega\kappa} (x-y)^\kappa\cdot
\delta\left[n.(x-y)\right] \left[{\cal L}_I^{(2)}(x),{\cal L}_I^{(1)}(y)\right]
\nonumber\\ 
 &\equiv& P_{\omega\kappa}\frac{\delta}{\delta n_\kappa}S^{(2)}_{c.t.}\ .
\label{app:EX.29}\end{eqnarray}
Then, a term  $P_{\rho\sigma}\partial^\sigma\delta^4(x-y)$ in the commutator of the
interaction lagrangians gives
\begin{eqnarray*}
&& P_{\omega\kappa}(x-y)^\kappa \ldots P_{\rho\sigma}\partial^\sigma\delta^4(x-y)
 \stackrel{p.i.}{\Rightarrow} -P_{\omega\kappa} g^{\kappa\sigma} \ldots
 P_{\rho\sigma}\delta^4(x-y) \nonumber\\ &\Rightarrow&
-P_{\omega\kappa}P^\kappa_{\,\, \rho} \ldots \delta^4(x-y) = 
-P_{\omega\rho} \ldots \delta^4(x-y)\ .
\end{eqnarray*}
Therefore, we have the recipe for dealing with one derivative of the $\delta^4(x-y)$-
function:
\begin{equation}
 P_{\rho\sigma}\partial^\sigma \Rightarrow - P_{\omega\rho}\ .
\label{app:EX.30}\end{equation}
Now, the strategy is to modify the T-product by including $\tau$-terms, such as
to cancel the so called 'contact terms'. 
Let us define 
\begin{eqnarray}
 P_{\omega\kappa}\frac{\delta}{\delta n_\kappa}\tau^{(2)} &\Leftarrow& 
+\frac{1}{2!}\int d^4x\ \int d^4y\ P_{\omega\kappa} (x-y)^\kappa\cdot
\delta\left[n.(x-y)\right] \left[{\cal L}_I^{(2)}(x),{\cal L}_I^{(1)}(y)\right]\ .
\label{app:EX.31}\end{eqnarray}
So, specializing to $\pi N$, the $\tau$-terms have to satisfy
\begin{equation}
 P_{\omega\kappa}\frac{\delta}{\delta n_\kappa}\tau(p',q';p,q) = 
  -P_{\omega\kappa}\frac{\delta}{\delta n_\kappa}S^{(2)}(p',q';p,q)
\label{app:EX.32}\end{equation}
and so they can be obtained from (\ref{app:EX.28}), using replacements like in
(\ref{app:EX.30}).
Moreover, the factor $1/2!$ can be dropped because there is an identical contribution
form the '12-commutator'.

We split the contributions to the $\tau(x-y;n)$-function into two parts:
\begin{equation}
 \tau(x-y;n) = \tau_1(x-y;n) + \tau_2(x-y;n)\ , 
\label{app:EX.33}\end{equation}
where now the $\tau_{1,2}$-functions satisfy
\begin{eqnarray}
 P_{\omega\sigma}\frac{\partial}{\partial_\sigma}\tau_1(x-y;n) &=& 
+\frac{i}{3} M_\Delta
g^2\epsilon^{\mu\nu\alpha\beta}\epsilon^{\rho\lambda\gamma\delta}
\cdot\left(\partial_\beta\phi(x)\ \partial_\delta\phi(y)\right)\ .
\cdot\nonumber\\ && \times\left[
\bar{\psi}(x)\gamma_\alpha \left(g_{\lambda\nu}+\gamma_\lambda\gamma_\nu\right)
\gamma_\gamma \psi_(y)\right]
\cdot\nonumber\\ && 
\times\left(n_\mu P_{\rho\omega}+ n_\rho P_{\mu\omega}\right)\delta^4(x-y)\ ,      
\label{app:EX.34}\end{eqnarray}
and 
\begin{eqnarray}
 P_{\omega\sigma}\frac{\partial}{\partial_\sigma}\tau_2(x-y;n) &=&
+\frac{1}{3} g^2\epsilon^{\mu\nu\alpha\beta}\epsilon^{\rho\lambda\gamma\delta}
\cdot\left(\partial_\beta\phi(x)\ \partial_\delta\phi(y)\right)
\cdot\nonumber\\ &&\times\left[
\bar{\psi}(x)\gamma_\alpha \left(g_{\lambda\nu}+\gamma_\lambda\gamma_\nu\right)
\gamma^\epsilon\gamma_\gamma \psi_(y)\right]
\cdot\nonumber\\ && \times\left\{
\vphantom{\frac{A}{A}}
\left(P_{\mu\omega} P_{\rho\sigma} n_\epsilon + P_{\epsilon\omega}P_{\rho\sigma} n_\mu
 + P_{\epsilon\omega} P_{\mu\sigma} n_\rho \right) \partial^\sigma\right.
\nonumber\\ && \left.
 +\left(P_{\mu\kappa} P_{\rho\omega} n_\epsilon + P_{\epsilon\kappa}P_{\rho\omega} n_\mu
 + P_{\epsilon\kappa} P_{\mu\omega} n_\rho \right) \partial^\kappa
 \right.\nonumber\\ && \left.
 -n_\mu n_\rho n_\epsilon\left(P_{\omega\sigma}\partial^\sigma 
 + P_{\omega\kappa}\partial^\kappa
\vphantom{\frac{A}{A}}\right) \right\} \delta^4(x-y)
\nonumber\\ &=&
+\frac{1}{3} g^2\epsilon^{\mu\nu\alpha\beta}\epsilon^{\rho\lambda\gamma\delta}
\cdot\left(\partial_\beta\phi(x)\ \partial_\delta\phi(y)\right)
\cdot\nonumber\\ && \times\left[
\bar{\psi}(x)\gamma_\alpha \left(g_{\lambda\nu}+\gamma_\lambda\gamma_\nu\right)
\gamma^\epsilon\gamma_\gamma \psi_(y)\right]
\cdot\nonumber\\ && \times\left\{
\vphantom{\frac{A}{A}}
 \left(P_{\epsilon\omega}P_{\rho\sigma}+P_{\epsilon\sigma}P_{\rho\omega}\right) n_\mu 
 +\left(P_{\epsilon\omega}P_{\mu\sigma}+P_{\epsilon\sigma}P_{\mu\omega}\right) n_\rho     
 \right.\nonumber\\ && \left.
 +\left(P_{\mu\omega} P_{\rho\sigma} + P_{\mu\sigma} P_{\rho\omega}
 -2 n_\mu n_\rho P_{\omega\sigma}\right) n_\epsilon 
\vphantom{\frac{A}{A}}
 \right\}\ \partial^\sigma \delta^4(x-y)\ .
\label{app:EX.35}\end{eqnarray}

\noindent Next, for the contribution to the $\pi N$-matrix elements
we evaluate the derivative $\partial^\sigma$. There are two contributions:\\

\noindent (i) Factor 
\begin{eqnarray*}
&& \partial^\sigma\left[\bar{\psi}(x)\psi(y)\right] \rightarrow 
\exp\frac{i}{2}\left[(p'+p)\cdot(x-y)+(p'-p)\cdot(x+y)\right] \nonumber\\
&& \Rightarrow -\frac{i}{2}(p'+p)^\sigma\ , \nonumber\\
&& \left(\partial_\beta\phi(x)\ \partial_\delta\phi(y)\right) \rightarrow 
 \left(q_\beta q'_\delta + q'_\beta q_\delta\right)\ , 
\end{eqnarray*}
\noindent (ii) Factor 
\begin{eqnarray*}
&& \partial^\sigma\left(\partial_\beta\phi(x)\ \partial_\delta\phi(y)\right) \rightarrow 
 -\frac{i}{2}(q'+q)^\sigma\left(q_\beta q'_\delta - q'_\beta q_\delta\right)\ . 
\end{eqnarray*}
We now write 
\begin{eqnarray}
\langle p'q'|\tau^{(2)}_{c.t.}|p,q\rangle &=& (2\pi)^4 \delta(p'+q'-p-q)\ 
 \bar{\tau}(p',q';p,q)\ \ {\rm with} \nonumber\\ 
 \bar{\tau}(p',q';p,q) &=& \bar{\tau}_1(p',q';p,q) + \bar{\tau}_2(p',q';p,q)\ ,        
\label{app:EX.36a}\end{eqnarray}
and find from (\ref{app:EX.34}) and (\ref{app:EX.35}) that 
\begin{eqnarray}
P_{\omega\sigma}\frac{\delta}{\delta n_\sigma}\bar{\tau}_1(p',q';p,q) &=&
+\frac{2i}{3} M_\Delta g^2\epsilon^{\mu\nu\alpha\beta}\epsilon^{\rho\lambda\gamma\delta}
\left(q_\beta q'_\delta + q'_\beta q_\delta\right)
\cdot\nonumber\\ && \times\left[
\bar{u}(p')\gamma_\alpha \left(g_{\lambda\nu}+\gamma_\lambda\gamma_\nu\right)
\gamma_\gamma u(p)\right] \cdot\nonumber\\ && 
\times\left(n_\mu P_{\rho\omega}+ n_\rho P_{\mu\omega}\right)\ ,      
\label{app:EX.36b}\end{eqnarray}
and
\begin{eqnarray}
 P_{\omega\sigma}\frac{\delta}{\delta n_\sigma}\bar{\tau}_2(p',q';p,q) &=&
-\frac{i}{3} g^2\epsilon^{\mu\nu\alpha\beta}\epsilon^{\rho\lambda\gamma\delta}
\cdot\nonumber\\ && \times \left[
\left(q_\beta q'_\delta + q'_\beta q_\delta\right) Q^\sigma
-\left(q_\beta q'_\delta - q'_\beta q_\delta\right) P^\sigma\right]
\cdot\nonumber\\ && \times\left[
\bar{u}(p')\gamma_\alpha \left(g_{\lambda\nu}+\gamma_\lambda\gamma_\nu\right)
\gamma^\epsilon\gamma_\gamma u(p)\right]  
\cdot\nonumber\\ && \times\left\{ \vphantom{\frac{A}{A}}
 \left(P_{\epsilon\omega}P_{\rho\sigma}+P_{\epsilon\sigma}P_{\rho\omega}\right) n_\mu 
 +\left(P_{\epsilon\omega}P_{\mu\sigma}+P_{\epsilon\sigma}P_{\mu\omega}\right) n_\rho     
 \right.\nonumber\\ && \left.
 +\left(P_{\mu\omega} P_{\rho\sigma} + P_{\mu\sigma} P_{\rho\omega}
 -2n_\mu n_\rho P_{\omega\sigma}\right) n_\epsilon 
\vphantom{\frac{A}{A}} \right\}\ \partial^\sigma \delta^4(x-y)\ , 
\label{app:EX.37}\end{eqnarray}
where $P \equiv (p'+p)/2$ and $Q \equiv (q'+q)/2$.\\

\noindent (i) Using FORM \cite{Vermaseren} we obtain for $\bar{\tau}_1$:
\begin{eqnarray}
&& P_{\mu\nu}\frac{\delta}{\delta n_\nu}\bar{\tau}_1(p',q';p,q) =
 +\frac{i}{3} M_\Delta g^2\left[\vphantom{\frac{A}{A}} 
8 (n\cdot q)(n\cdot q')\ n_\mu -4(n\cdot q') q_\mu -4(n\cdot q) q'_\mu\right]\ .
\label{app:EX.38}\end{eqnarray}
To solve this we consider the equations 
\begin{eqnarray*}
 P_{\mu\nu} a^\nu &\equiv & 8 (n\cdot q)(n\cdot q')\ n_\mu -4(n\cdot q') q_\mu 
 -4(n\cdot q) q'_\mu\ \ ,\ \ {\rm where} \nonumber\\
 a_\mu &=& a_1\ q_\mu + a_2\ q'_\mu + a_3\ n_\mu\ , 
\end{eqnarray*}
which has as solution
\begin{eqnarray*}
&& a_1 = -4 (n\cdot q')\ \ ,\ \ a_2 = -4 (n\cdot q)\ \ ,\ \ a_3= {\it undetermined}\ .
\end{eqnarray*}
So,
\begin{eqnarray*}
a_\mu &=& -4(n\cdot q') q_\mu -4(n\cdot q) q'_\mu + a_3 n_\mu\ ,     
\end{eqnarray*}
which gives as solution for $\bar{\tau}_1$
\begin{eqnarray}
\bar{\tau}_1(p',q';p,q) & = & -4i (n\cdot q') (n\cdot q)
 \times\frac{1}{3}g^2 M_\Delta\ .
\label{app:EX.39}\end{eqnarray}

\noindent (ii) Using FORM \cite{Vermaseren} we obtain for $\bar{\tau}_2$:
\begin{eqnarray}
 P_{\mu\nu}\frac{\delta}{\delta n_\nu}\bar{\tau}_2(p',q';p,q) &=&
 +\frac{i}{3} g^2\left\{ \left[ 8 \mbox{$n \hspace{-0.45em}/$}\left( 
\vphantom{\frac{A}{A}}
(n\cdot q)(q'\cdot P) + (n\cdot q')(q\cdot P)
 \right.\right.\right.\nonumber\\ && \left.\left.\left.\vphantom{\frac{A}{A}}
-4(n\cdot q)(n\cdot q')(n\cdot P)\right) + \mbox{$P \hspace{-0.45em}/$}
(n\cdot q')(n\cdot q) \right] n_\mu \right. \nonumber\\ && \left.
 + \left[\mbox{$n \hspace{-0.45em}/$} \left(\vphantom{\frac{A}{A}} 
 8 (n\cdot q')(n\cdot P) -4 (q'\cdot P)\right) - 
\mbox{$P \hspace{-0.45em}/$}(n\cdot q')\right] q_\mu
\right.\nonumber\\ && \left.
 + \left[\mbox{$n \hspace{-0.45em}/$} \left(\vphantom{\frac{A}{A}} 
 8 (n\cdot q)(n\cdot P) -4 (q\cdot P)\right) - 
\mbox{$P \hspace{-0.45em}/$}(n\cdot q)\right] q'_\mu
 + 8\mbox{$n \hspace{-0.45em}/$}(n\cdot q')(n\cdot q)\ P_\mu  
\right.\nonumber\\ && \left. 
 + \left[\vphantom{\frac{A}{A}} 8(n\cdot q')(n\cdot q)(n\cdot P) 
 - 4(n\cdot q')(q\cdot P) - 4(n\cdot q)(q'\cdot P)\right] \gamma_\mu\right\}
\label{app:EX.40}\end{eqnarray}
Repeating the procedure above, we write
\begin{eqnarray*}
&& P_{\mu\nu}\delta\bar{\tau}_2/\delta n_\mu \propto P_{\mu\nu} b^\nu \equiv \left\{
\vphantom{\frac{A}{A}} \ldots \right\}\ \ {\rm with} \nonumber\\ 
&& b_\mu = b_1\ q_\mu + b_2\ q'_\mu + b_3\ P_\mu + b_4 \gamma_\mu + b_5 n_\mu\ .
\end{eqnarray*}
Then, we obtain the coefficients 
\begin{eqnarray*}
b_1 &=&  8\mbox{$n \hspace{-0.45em}/$} (n\cdot q')(n\cdot P)   
        -4\mbox{$n \hspace{-0.45em}/$} (q'\cdot P)   
        -4\mbox{$P \hspace{-0.45em}/$} (n\cdot q')\ , \nonumber\\
b_2 &=&  8\mbox{$n \hspace{-0.45em}/$} (n\cdot q)(n\cdot P)   
        -4\mbox{$n \hspace{-0.45em}/$} (q\cdot P)   
        -4\mbox{$P \hspace{-0.45em}/$} (n\cdot q')\ , \nonumber\\
b_3 &=&  8\mbox{$n \hspace{-0.45em}/$} (n\cdot q)(n\cdot q')\ , \nonumber\\
b_4 &=&  8 (n\cdot q)(n\cdot q')(\cdot P) -4(n\cdot q)(q'\cdot P)
         -4(n\cdot q')(q\cdot P)\ , \nonumber\\
b_5 &=&  {\it undetermined}\ .                                      
\end{eqnarray*}
From this we get as a solution for $\bar{\tau}_2$:
\begin{eqnarray}
\bar{\tau}_2(p',q';p,q) &=& +\frac{i}{3}g^2\left[
\vphantom{\frac{A}{A}}
  -4\mbox{$P \hspace{-0.45em}/$} (n\cdot q')(n\cdot q)   
  -4\mbox{$n \hspace{-0.45em}/$} (n\cdot q)(P\cdot q')   
  -4\mbox{$n \hspace{-0.45em}/$} (n\cdot q')(P\cdot q)   
 \right.\nonumber\\ && 
 \left.\vphantom{\frac{A}{A}}
  +8\mbox{$n \hspace{-0.45em}/$} (n\cdot q')(n\cdot q)(n\cdot P)\right]\ .
\label{app:EX.41}\end{eqnarray}
Again, these $\tau$-functions cancel the non-covariant, i.c. the 'frame-dependent',
terms produced by the T-product in second order. Also, they are the right 
corrections to the Kadyshevsky amplitudes such as to give agreement with 
the Feynman-amplitudes when $\kappa=\kappa'=0$.



\section{Discussion and Conclusions}                        
\label{sec:Z}
We have shown that a functional integral formulation can be
formulated for the Kadyshevky theory using a second quantization texchnique, 
which leads to a path-integral,
Schwinger-Symanzik equations, Kadyshevsky reduction formulas.
Therefore, the Kadyshevky formulation allows one to use 
all techniques which can be utilized in studies of field theories like those
in the Feynman formulation. For example, non-abelian theories like QCD and
the Electro-weak theories can be studied in the Kadyshevsky form.

Also, we have shown that frame independence in the Kadyshevsky formalism can be 
achieved by following the approach of Gross-Jackiw \cite{Gross69}. 
Application,
The analogon of the $T^*$-product is used to demonstrate the Lorentz-invariance of the S-matrix
also for interactions that contain derivatives. Application 
to the pion-nucleon amplitudes shows that on-energy-shell, i.e. $\kappa=\kappa'=0$
leads to amplitudes identical to those with the Feynman-formalism, also
for amplitudes where the $\Delta_{33}$-resonance is involved.


 \begin{figure}[hhhhtb]
 \resizebox{11.25cm}{!}
 {\includegraphics[width=10cm,height=8cm]{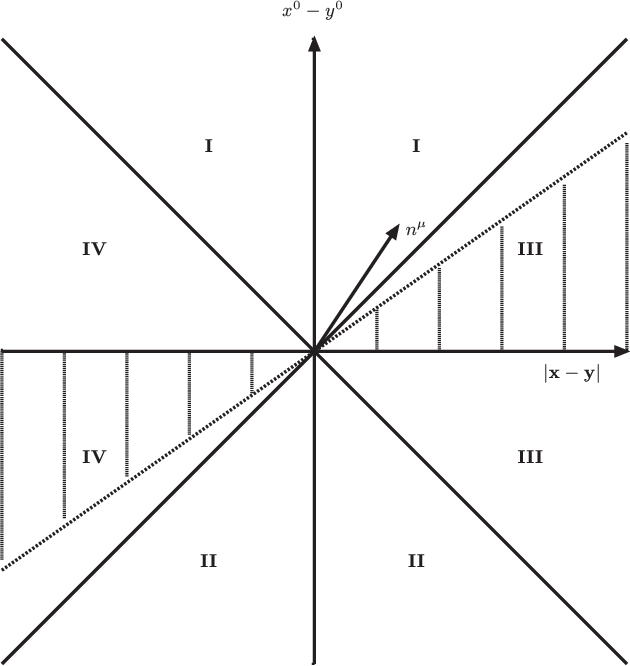}}
 \caption{\sl Minkowski-plane. The dashed lines mark the points $n\cdot(x-y)=0$. 
 In the regions I and II $(x-y)^2 > 0$, and in the regions III and IV $(x-y)^2 < 0$.
 }                         
  \label{fig:minkowski} 
  \end{figure}                     

\appendix

\section{Kadyshevsky-rules in Momentum-space}                              
\label{app:A}   
The invariant amplitude $-M_{\kappa',\kappa}$ \ref{app:A.8}) is computed by drawing all
connected Feynman graphs for the considered process. The amplitude 
\[               
-(2\pi)^4\delta(\sum_i p_{i,out}+\kappa' n -\sum_i p_{i,in}-\kappa n) 
M_{\kappa',\kappa}(G)  
\]    
corresponding to graph $G$ is built up by associating
factors with the elements of the graph, which we list below:\\

\noindent I. Those factors, independent of the 
specific details of the interactions, are given by the following rules:\\

\noindent\hspace{1cm} 1. Draw the Feynman graph G. Arbitrarily number its 
vertices and orient each internal particle line from the vertex with the smaller
number to the vertex with the larger number, assigning to it a 4-momentum
$p$.

\noindent\hspace{1cm} 2. Connect with dotted lines the first vertex with the 
second, the second with the third, etc. Orient them in the direction of
increasing numbers and assign to them a 4-momentum $\kappa_s n$, where
$s=1,2, \dots, n-1$ is the number of the vertex which a given dotted line
leaves. Attach to the first vertex an incoming external dotted line with
4-momentum $\kappa_i n$, and to the last vertex $n$ an outgoing external
dotted line with 4-momentum $\kappa_f n$.

\noindent\hspace{1cm} 3. For incoming (outgoing) boson and fermion lines:
identical to the rules for Feynman graphs \cite{BD65}. 

\noindent\hspace{1cm} 4. For each internal dotted line with momentum 
$\kappa n$ a factor
\begin{equation}
 G_0(\kappa) = -\frac{1}{\kappa+i\epsilon}\ .
\label{app:A.9}\end{equation}

\noindent\hspace{1cm} 5. For each internal boson line with momentum $q$ 
a factor
\begin{equation}
 \Delta^{(+)}(q) = \theta(q_0) \delta(q^2-\mu^2)\ .      
\label{app:A.10}\end{equation}

\noindent\hspace{1cm} 6. For each internal fermion line with momentum $p$ 
and {\it positive energy} a factor
\begin{equation}
 S^{(+)}_{\beta\alpha}(p) =          
 \left(\mbox{$p \hspace{-0.45em}/$} + m\right)_{\beta\alpha}
 \theta(p_0) \delta(p^2-m^2)\ .
\label{app:A.11}\end{equation}
\hspace{1.5cm} For each internal fermion line with momentum $p$ 
and {\it negative energy} a factor
\begin{equation}
 S^{(-)}_{\beta\alpha}(p) =          
 \left(\mbox{$p \hspace{-0.45em}/$} - m\right)_{\beta\alpha}
 \theta(p_0) \delta(p^2-m^2)\ .
\label{app:A.12}\end{equation}

\noindent\hspace{1cm} 7. For each internal photon line, using the Feynman
gauge, a factor
\begin{equation}
 D^{(+)}(q)_{\mu\nu} = -g_{\mu\nu} \theta(q_0) \delta(q^2)\ .      
\label{app:A.13}\end{equation}

\noindent\hspace{1cm} 8a. For each vertex, number $s$, a factor 
\begin{equation}
  (2\pi)^4 \delta^4\left(\sum_{i} p_{i,out}+\kappa_{s+1}- \sum_{i} p_{i,in}
 -\kappa_s\right)\ ,
\label{app:A.14}\end{equation}
\hspace{1.5cm} where $p_{i,out}$ and $p_{i,in}$ are the outgoing respectively the
incoming momenta \\
\hspace{1.5cm} at the vertex with number $s$.

\noindent\hspace{1cm} 8b. Integrate over each internal particle line  
, momentum $l$:
 $\int d^4l/(2\pi)^3$.                                      

\noindent\hspace{1cm} 9. 
Integrate over each internal quasi-particle (dotted) line \\
\hspace{1.5cm} with momentum $\kappa_s n$:         
$\int_{-\infty}^{+\infty} d\kappa_s/(2\pi)$.                        

\noindent\hspace{1cm} 10. {\it Not} a factor $-1$ for each closed loop.          

\noindent\hspace{1cm} 11. A factor $-1$ between graphs which differ only 
by an interchange of two-external fermions. This not only for the interchange 
of identical fermions in the final state, but also the interchange of e.g. an 
initial fermion and a similar anti-fermion in the final state.

\noindent\hspace{1cm} 12. Repeat the operations (1)-(11) for all $n!$    
numbering's of the vertices of the given Feynman graph and sum. \\

\noindent II. Those factors coming from the structure and type of vertices 
are, given for each vertex by the matrix element 
$\langle \ldots | {\cal L}_I(0)| \ldots\rangle$. Therefore, they are,
apart from a factor $(-i)$, identical to that given in \cite{BD65}, 
appendices B. 

\section{Kadyshevsky $\widetilde{T}$-products and Wick's Theorem}          
\label{app:L}   

In this appendix we treat the free scalar fields.
The Kadyshevsky $\widetilde{T}$-product we define as 
\begin{equation}
\widetilde{T}\left[\phi(x) \phi(y)\right] = \theta[n\cdot(x-y)]\ \phi(x) \phi(y)
 + \theta[n\cdot(y-x)]\ \phi(y) \phi(x)\ .
\label{app:L.1}\end{equation}
We have the obvious identity
\begin{equation}
\widetilde{T}\left[\phi(x) \phi(y)\right] = \phi(y) \phi(x)
 + \theta[n\cdot(x-y)]\ \left[\phi(x), \phi(y)\right]\ ,
\label{app:L.2}\end{equation}
and a similar expression for the ordinary T-product $T[\phi(x) \phi(y)]$. For the difference
we obtain
\begin{eqnarray}
&& \widetilde{T}\left[\phi(x) \phi(y)\right] - T[\phi(x) \phi(y)] = 
 + \left\{\theta[n\cdot(x-y)] - \theta(x^0-y^0)\right\}\ \left[\phi(x),\phi(y)\right] = 
 \nonumber\\ && 
 \left\{\theta[n\cdot(x-y)] - \theta(x^0-y^0)\right\}\ \Delta(x-y; m^2)\ .               
\label{app:L.3}\end{eqnarray}
In Fig.~\ref{fig:minkowski} the areas I and II are respectively the forward and backward
light-cone, where $(x-y)^2 > 0$. In the areas III and IV the distances are space-like,
i.e. $(x-y)^2 <0$.
Now, as seen in Fig.~\ref{fig:minkowski} in the arched area                       
light-cone,
\begin{eqnarray}
 \left\{\theta[n\cdot(x-y)] - \theta(x^0-y^0)\right\} \neq 0\ \ ,        
\label{app:L.4}\end{eqnarray}
which is outside the light-cone,
where $(x-y)^2 <0$. But, for this region $\Delta(x-y;m^2)=0$. Therefore, 
\begin{equation}
\widetilde{T}\left[\phi(x) \phi(y)\right] = T\left[\phi(x) \phi(y)\right]\ . 
\label{app:L.5}\end{equation}
For the interaction Lagrangian micro-causality reads
\begin{equation}
 \left[ {\cal L}_I(x), {\cal L}_I(y) \right] = 0\ \ ,\ \ {\rm for}\ \ (x-y)^2<0\ .
\label{app:L.6}\end{equation}
Therefore, from the above we can infer immediately that
\begin{equation}
\widetilde{T}\left[{\cal L}_I(x) {\cal L}_I(y)\right] = 
T\left[{\cal L}_I(x) {\cal L}_I(y)\right]\ , 
\label{app:L.7}\end{equation}
a result that can be generalized immediately to a T-product of any number of interaction
Lagrangians, i.e. 
\begin{equation}
\widetilde{T}\left[{\cal L}_I(x_1)\ \ldots\  {\cal L}_I(x_n)\right] = 
T\left[{\cal L}_I(x_1)\ \ldots\  {\cal L}_I(x_n)\right]\ . 
\label{app:L.8}\end{equation}
From this one can conclude to the complete equivalence of (\ref{app:A.1}) and 
(\ref{app:A.3}). (Q.E.D.)\\

\noindent Wick's Theorem in the case of the Kadyshevsky $\widetilde{T}$-product 
reads (for n=even)
\begin{eqnarray}
&& \widetilde{T}\left(\phi(x_1) \ldots \phi(x_n) \right) = : \phi(x_1) \ldots \phi(x_n): 
\nonumber\\ && +\ \left[\langle 0|\widetilde{T}\left(\phi(x_1)\phi(x_2)\right)|0\rangle\
:\phi(x_3)\ \ldots\ \phi(x_n): + {\rm permutations} \right] 
\nonumber\\ && +\ \left[\langle 0|\widetilde{T}\left(\phi(x_1)\phi(x_2)\right)|0\rangle\
\langle 0|\widetilde{T}\left(\phi(x_3)\phi(x_4)\right)|0\rangle\
:\phi(x_5)\ \ldots\ \phi(x_n): 
\right. \nonumber\\ && \left. \hspace{10cm} + {\rm permutations} \right] 
\nonumber\\ && +\ \ldots \nonumber\\ && 
 +\ \left[\langle 0|\widetilde{T}\left(\phi(x_1)\phi(x_2)\right)|0\rangle\ \ldots\ 
          \langle 0|\widetilde{T}\left(\phi(x_{n-1})\phi(x_n)\right)|0\rangle       
\right. \nonumber\\ && \left. \hspace{10cm} + {\rm permutations} \right]\ ,
\label{app:L.9}\end{eqnarray}
and a somewhat different form for n=uneven.
This is the same as for the ordinary T-product, see \cite{BD65}, section (17.4). 
To prove this, we consider first the case n=2. Then, in terms of the positive and 
negative frequency parts of the field, we have
\begin{eqnarray}
\widetilde{T}\left[\phi(x) \phi(y)\right] &=& : \phi(x) \phi(y): + 
  \theta[n\cdot(x-y)]\ \langle 0| \phi^{(+)}(x) \phi^{(-)}(y)|0\rangle +
\nonumber\\ &&
  \theta[n\cdot(y-x)]\ \langle 0| \phi^{(+)}(y) \phi^{(-)}(x)|0\rangle \nonumber\\ 
 &=& : \phi(x) \phi(y): + 
\langle 0| \widetilde{T}\left[\phi(x) \phi(y)\right] |0\rangle\ ,
\label{app:L.10}\end{eqnarray}
where we used the fact that 
 $\langle 0| \phi^{(+)}(x) \phi^{(-)}(y)|0\rangle = \langle 0| \phi(x) \phi(y)|0\rangle $
etc. (Q.E.D.) \\
\noindent For the general case we could follow the usual proof, see \cite{BD65}, but 
we prefer to exploit here Lorentz-transformation properties. Thereto, we consider the
Lorentz-transformation $U(a)$ of the scalar field
\begin{equation}
 U(a) \phi(x) U^{-1}(a) = \phi(x')\ \ {\rm where}\ \ x' = a x\ , 
\label{app:L.11}\end{equation}
leaving the vacuum invariant, i.e. $U(a) |0\rangle = |0\rangle$. 
We now take $a$ such that $n^\mu= a^\mu_{\hspace{1mm} \nu} \hat{n}^\nu$, where
$\hat{n}^\mu = (1,{\bf 0})$. Also, $\phi(x) = U(a) \phi(\hat{x}) U^{-1}(a)$ and 
$\theta[n\cdot(x-y)] = \theta(\hat{x}^0-\hat{y}^0)$. 
Then, 
\begin{equation}
\widetilde{T}\left(\phi(x_1) \ldots \phi(x_n) \right) =  
 U(a) \left\{ T\left(\phi(\hat{x}_1)\ \ldots\ \phi(\hat{x}_n) \right)\right\} U^{-1}(a)\ ,
\label{app:L.12}\end{equation}
and Wick's Theorem for the Kadyshevsky $\widetilde{T}$-product follows 
linea recta from that for the ordinary T-product. (Q.E.D)\\

The equivalence of the S-matrix expressions (\ref{app:A.1}) and (\ref{app:A.3})
can now be seen again also in the following way.
From the Wick-expansion we see that the Kadyshevsky $\widetilde{T}$-product
for any number of fields is equivalent to the ordinary T-product, if the Kadyshevsky
$\widetilde{T}$-product for two fields is the same. 
The latter has been demonstrated explicitly above 
in (\ref{app:L.3})-(\ref{app:L.5}). (Q.E.D.)


\end{document}